\documentclass[]{imsart}

\usepackage[margin=1in]{geometry}
\usepackage[toc]{appendix}
\usepackage{graphicx}
\RequirePackage[OT1]{fontenc}
\RequirePackage{natbib}
\usepackage{adjustbox}
\usepackage{amsmath,amssymb,amsthm,bm,latexsym}
\usepackage{graphicx,psfrag,epsf}
\usepackage{epigraph,xcolor}
\usepackage{enumerate}
\usepackage[hyphens]{url}
\usepackage{booktabs}
\usepackage{multirow}
\usepackage{hyperref}
\usepackage{float}

\usepackage{algpseudocode}
\usepackage[linesnumbered,ruled]{algorithm2e}
\hypersetup{colorlinks=true,citecolor=blue,urlcolor=blue,
	pdfpagemode=FullScreen,linktocpage=true}
\usepackage{cleveref}

\allowdisplaybreaks

\startlocaldefs

\def\R{\mathbb{R}}

\newcommand{\abs}[1]{\left\lvert #1 \right\rvert}
\newcommand{\norm}[1]{\left\| #1 \right\|}

\newcommand{\corr}{\mathrm{Corr}}
\newcommand{\I}{\mathbf{I}}

\renewcommand{\S}{\mathcal{S}}

\newtheoremstyle{general}
{3mm} 
{3mm} 
{\it} 
{} 
{\bfseries} 
{.} 
{.5em} 
{} 

\theoremstyle{general}



\endlocaldefs

\begin{document}

\begin{frontmatter}

\title{A Bayesian approach to identify changepoints in spatio-temporal ordered categorical data: An application to COVID-19 data}
\runtitle{ Identifying changepoints in spatio-temporal ordered categorical data}

\begin{aug}
\author[A]{\fnms{Siddharth} \snm{Rawat}\ead[label=e1]{siddharth.rawat19@iimb.ac.in}},
\author[B]{\fnms{Abe} \snm{Durrant}},
\author[B]{\fnms{Adam} \snm{Simpson}},
\author[B]{\fnms{Grant} \snm{Nielson}},
\author[B]{\fnms{Candace} \snm{Berrett}}
\and
\author[A]{\fnms{Soudeep} \snm{Deb}},
  
\address[A]{Indian Institute of Management Bangalore \\ Bannerghatta Main Rd, Bangalore, KA 560076, India. \\ Corresponding author: S. Rawat, Email: \printead{e1}}

\address[B]{Department of Statistics, Brigham Young University, Provo, UT, USA}

\runauthor{Rawat et al. }
\end{aug}

\begin{abstract}
Although there is substantial literature on identifying structural changes for continuous spatio-temporal processes, the same is not true for categorical spatio-temporal data. This work bridges that gap and proposes a novel spatio-temporal model to identify changepoints in ordered categorical data. The model leverages an additive mean structure with separable Gaussian space-time processes for the latent variable. Our proposed methodology can detect significant changes in the mean structure as well as in the spatio-temporal covariance structures. We implement the model through a Bayesian framework that gives a computational edge over conventional approaches. From an application perspective, our approach's capability to handle ordinal categorical data provides an added advantage in real applications. This is illustrated using county-wise COVID-19 data (converted to categories according to CDC guidelines) from the state of New York in the USA. Our model identifies three changepoints in the transmission levels of COVID-19, which are indeed aligned with the ``waves'' due to specific variants encountered during the pandemic. The findings also provide interesting insights into the effects of vaccination and the extent of spatial and temporal dependence in different phases of the pandemic.
\end{abstract}


\end{frontmatter}

\section{Introduction}
\label{sec:intro}

The changepoint problem has rich existing literature. The earliest work by \cite{chernoff1964estimating} analyzed a change in the mean structure for an independent and identically distributed (i.i.d.)  Gaussian distribution.  Ensuing work examined changepoints in the variance and structural changes for the non-Gaussian case \citep[see, for example,][]{zacks1983survey, krishnaiah198819}. Moving away from the i.i.d.\ Gaussian setting, changepoints are most often modeled or tested in time series data.  In this case, the concept of a changepoint implies a change in the temporal dependence after a specific point in time. The simplest case to account for a changepoint in time is to account for a change in the mean structure after the changepoint, while the autoregressive or dependence structure remains the same \citep[see][]{peter1991time}. \cite{taylor1994modeling} implemented a new class of models to account for stochastic volatility, and \cite{kim1998stochastic} analyzed similar research topics in the autoregressive conditional heteroskedasticity (ARCH) framework.

In fact, changepoint detection methods in time series data are abounding in extant literature. Our focus in this work is on spatio-temporal data sets; therefore, we provide a brief review of this literature here. \cite{majumdar2005spatio} proposed a model-based approach under a Bayesian framework and estimated a single changepoint in both the mean and dependence structures. \cite{yu2008multilevel} and \cite{xu2012multilevel} developed multi-level spatio-temporal dual changepoint models using conditional autoregressive (CAR) structure. They applied the model to examine the effect of alcohol outlets' control policy on assaultive violence rates. \cite{altieri2015changepoint} proposed a Bayesian changepoint model for spatio-temporal point processes and fit the model with the help of Integrated Nested Laplace Approximation (INLA) to detect multiple changepoints. \cite{altieri2016bayesian} extended the previous work to a Bayesian P-splines framework for examining earthquake point processes. \cite{fiedler2018multiple} studied a multiple-changepoints model in spatio-temporal seismic data, and implemented it through a Bayesian approach.  More recently, \cite{berrett2023} used a model selection framework for selecting the number of changes at any given location for Gaussian spatio-temporal data.

Although modeling changepoints for continuous data is most commonly encountered in practical problems, response variables are sometimes recorded as ordinal categories. A typical example is the case of survey responses. Moreover, even if the response variable is continuous, it often makes sense to convert it to an ordinal categorical variable due to the error in data collection and possible difficulty in count data modeling \citep[see][]{taylor1961aggregation, mullahy1997heterogeneity}. It is particularly important in the context of this paper, i.e., for COVID-19 case count data. It has been demonstrated that these case counts are noisy, generally under-reported, and often unreliable, primarily due to local test collection and upstream reporting errors \citep{dubrow2022local}. Hence, modeling it as an ordinal categorical response variable is more appropriate. In this work, we define the category of ``transmission level'' of COVID-19 from the county-level weekly new number of cases, following the guidelines from CDC \citep{cdc2022}. The transmission of such a disease depends on many non-measurable or imprecise quantities -- for example, the development of different virus variants and at-home testing availability -- creating changes in how true COVID-19 cases relate to measured quantities. We plan to examine the changing relationship of the transmission levels accounting for suitable covariates, including the important aspect of vaccination. 

One particular challenge in analyzing COVID-19 transmission is the spatial and temporal dependence patterns, as shown in \cite{rawat2021spatio}. This phenomenon complicates the covariance structure for the count data of the number of cases. For further challenges in this context, see \cite{bertozzi2020challenges}. Interestingly, although there is a loss of precision with the aforementioned formulation of the ordinal response variable, the spatio-temporal dependence remains, as we show through the exploratory analysis in \Cref{sec:data}. Thus, accounting for spatio-temporal dependence in the ordinal response variable is necessary.  

There is extensive literature for multinomial regression models of ordinal response variables, which are an extension of logistic and probit regression \citep[e.g.,][]{campbell1989classification, benoit2012multinomial, liang2020multinomial}. Bayesian models using a probit link use a latent variable to facilitate Gibbs sampling as shown by \cite{albert1993bayesian}. Bayesian estimation methodology and computational developments are also explored by \cite{tanner1987calculation}, who examined the calculation of posterior distribution using an approximate sampling method for inference. \cite{cowles1996accelerating} discussed Markov chain Monte Carlo (MCMC) approximation with an updated Gibbs sampling step to accelerate the convergence of the Bayesian estimation process \citep[see also,][]{imai2005, berrett2012, heiner2022}. Additionally, \cite{nandram1996reparameterizing} proposed the reparameterization of generalized linear models (GLM) to speed up the convergence of the Gibbs sampler. For the same purpose, \cite{liu1999parameter} proposed a parameter expanded data augmentation (PX-DA), which uses auxiliary variables to accelerate the convergence. \cite{chen2000unified} tackled the problem of estimating a GLM for correlated ordinal data with a scale mixture of a multivariate normal link function using a Bayesian framework with informative prior distributions. \cite{roy2007convergence} compared the performance and convergence rate of PX-DA  and the Bayesian data augmentation methodology proposed by \cite{albert1993bayesian}.

Although these methods are valuable, they have not proposed how to detect any changes in mean, variance, or correlation patterns in a categorical spatio-temporal data set. Thus, there is a dire need to develop a changepoint detection method that combines and expands on these modeling approaches to quantify relationships when there is spatial and temporal dependence. This is where we aim to fill the gap in the extant literature. Specifically, we propose a novel method to model changepoint(s) for spatio-temporal ordered categorical data using a Bayesian methodology. In the proposed model, a latent variable is used to model the categorical observations with both mean and variance-covariance structure shift after the changepoint, allowing the spatial and temporal associations to change. The model is implemented through a Bayesian framework, which gives it computational advantages over a classical approach. In this regard, we use the concepts of Gibbs sampling \citep{geman1984stochastic} and single variable slice sampling algorithm \citep{neal2003slice}.

Our methodology contributes to the literature in three exciting aspects. First, to the best of our knowledge, this is the first attempt to develop a method to detect changepoints in ordered categorical spatio-temporal data. Second, we provide a statistical way to identify the presence of a changepoint, and we extend the proposed approach in combination with a binary segmentation technique to a method that can detect multiple changepoints in the data. Third, our model employs an efficient estimation algorithm for all parameters in the model and alleviates some computational burden prevalent in similar spatio-temporal models for continuous response variables. We elaborate on this in \Cref{subsec:bayesian} below.

The rest of the paper is organized as follows. \Cref{sec:data} gives a brief description and outline of the data, along with the motivation of considering a spatio-temporal model. Our proposed method and related discussions are given in \Cref{sec:methods}. Application to the COVID-19 data from New York state, United States of America (USA), is presented in \Cref{sec:results}. We finish with a discussion and remarks in \Cref{sec:conclusion}. Some additional exploratory analyses are provided in \Cref{appendix-exp}, while thorough mathematical derivations of the Bayesian steps used in the method are provided in \Cref{appendix-post}.

\section{Data}
\label{sec:data}

The data used for the analysis are obtained from the GitHub repository maintained by \cite{datarepository}. The data is available at a county level. For this study, we concentrate on the data for all the counties from the state of New York. We look at the data at a weekly granularity, from 20th January 2020 to the week of 16th May 2022. This covers most of the COVID-19 ``waves'' in New York. Note that the data is prepared at a weekly level for two main reasons. First, this takes care of the problems with data reporting over weekends \citep[see,][]{ricon2020seven}. Second, the reduction in the size of the data from daily to weekly provides a significant computational advantage while not losing too much information on the data collection front. Thus,  the data set has a total of 7564 observations from 62 counties and 122 weeks. As mentioned in \Cref{sec:intro}, the weekly numbers are first converted into categorical data that reflect different transmission levels. We follow the CDC guidelines to define these levels, as provided in \Cref{tab:categ}.

\begin{table}[ht]
\caption{CDC guidelines used to define the ordered categories of COVID-19 transmission levels.}
\label{tab:categ}
\centering
\begin{tabular}{cc}
  \hline
New cases per 100,000 persons in a week
 & Category  \\ 
  \hline
  0 to 9.99 & 1 \\ 
  10 to 49.99 & 2 \\
  50 to 99.99 & 3 \\
  $\geqslant100$ & 4 \\
  \hline
\end{tabular}
\end{table}

\begin{figure}[!hbt]
\begin{center}
\includegraphics[width = 0.8\textwidth,keepaspectratio]{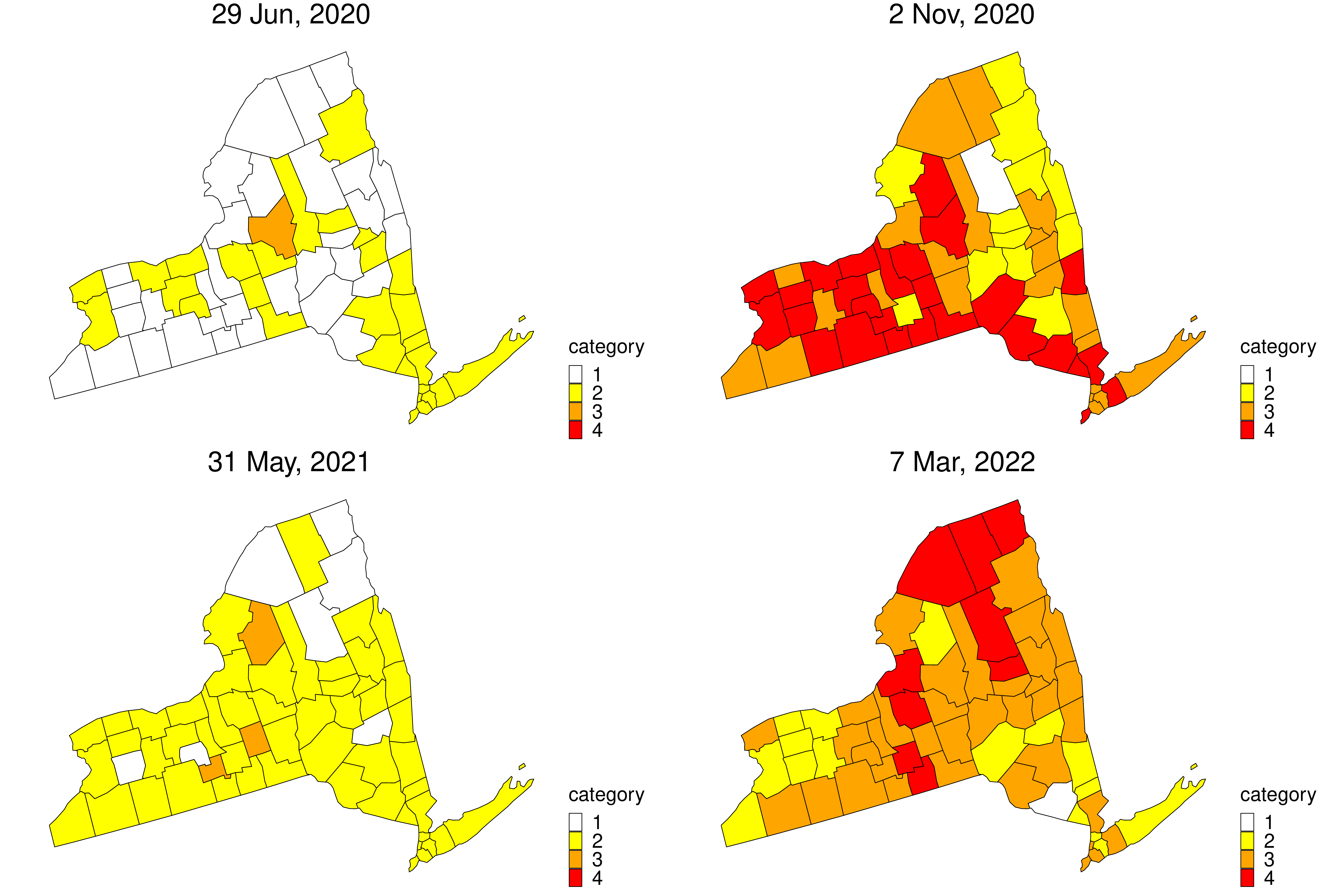}
\end{center}
\caption{COVID-19 transmission level category at four different time points for New York state counties.}
\label{fig:NY-map}
\end{figure}

As a first step of exploratory analysis, \Cref{fig:NY-map} shows the transmission level categories of COVID-19 in New York at four different weeks across the entire time period of the data set. At each point in time, there is clear spatial dependence for these categories. The maps show that counties with the same COVID-19 transmission levels tend to cluster, providing evidence of a need to account for or quantify spatial dependence in any modeling technique.

\begin{figure}[!hbt]
\begin{center}
\includegraphics[width=0.53\textwidth]{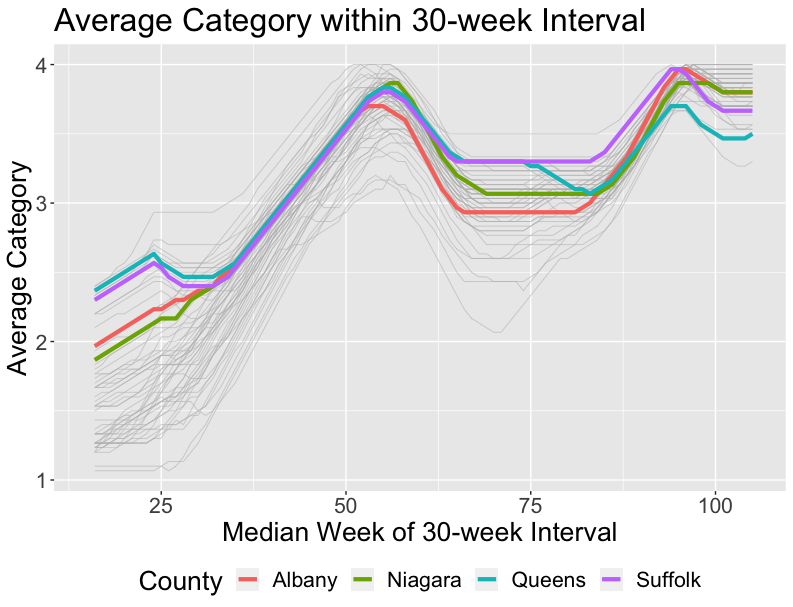}
\end{center}
\caption{Average transmission level for 30-week windows for all counties (gray) with four counties from across the state highlighted in the four colors.  }\label{fig:avgcat}
\end{figure}

We see similar dependence in the counties' transmission levels across time.  Because of the categorical nature of the response variable, it is easier to examine the similarities between counties in a smoother, less noisy plot.  Specifically, \Cref{fig:avgcat} shows the average transmission level for 30-week windows across the observed time period.  Each line represents a different county, with four counties -- Albany (red), Niagara (green), Queens (blue), and Suffolk (purple) -- highlighted in color for illustrative purposes. It is evident that the transmission levels are similar across counties for the entire time period.

We also account for two covariates in our model: the number of deaths that occurred in the county in the previous week (which reflects the severity of the virus) and the proportion of vaccinated people with the first dose (which reflects the extent of preventive measures). The rationale behind using these covariates is discussed in more detail in \Cref{subsec:app}. For the first one, in a similar line as \cite{rawat2021spatio}, we take the logarithmic transformation for the previous week's new death data and denote it by $\log d(s_i,t-1)$ for the $i^{th}$ county at the $t^{th}$ week. For the vaccination covariate, we define the vaccination prevalence $x(s_i, t)$ as 
\begin{equation}
\label{eqn:vaccine-prevalence}
x(s_i, t) = \frac{vc_{s_i,t}}{p_{s_i,t}},
\end{equation}
where $vc_{s_i,t}$ is the cumulative number of first doses of the COVID-19 vaccine administered in the $i^{th}$ county until the $t^{th}$ week, and $p_{s_i,t}$ is the corresponding population. In our model, for both covariates, we use standardized values such that the mean is 0 and the standard deviation is 1.

\begin{figure}[!hbt]
\begin{center}
\includegraphics[width=0.5\textwidth]{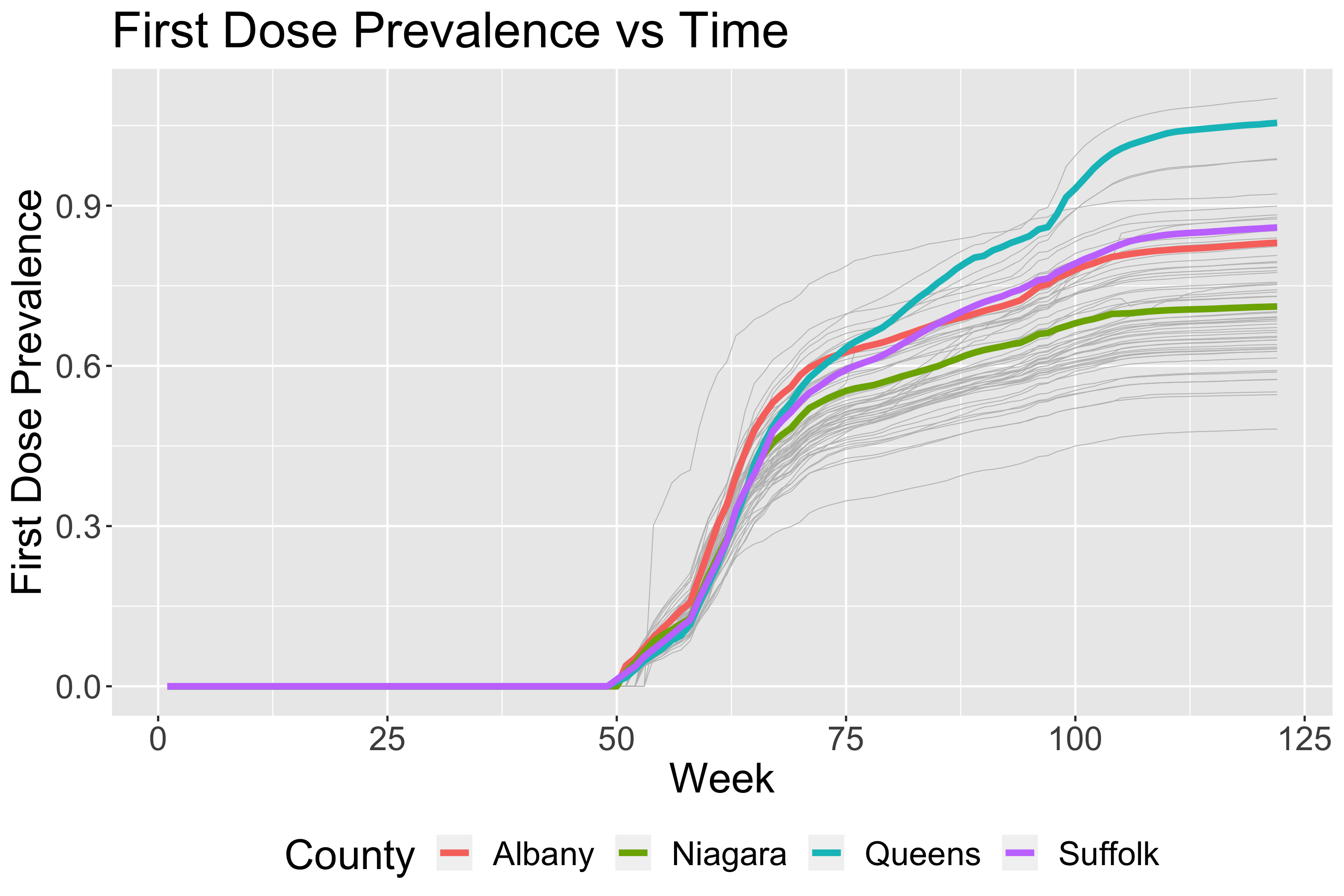}\\
(a)\\
\mbox{ }\\
\includegraphics[width=0.5\textwidth]{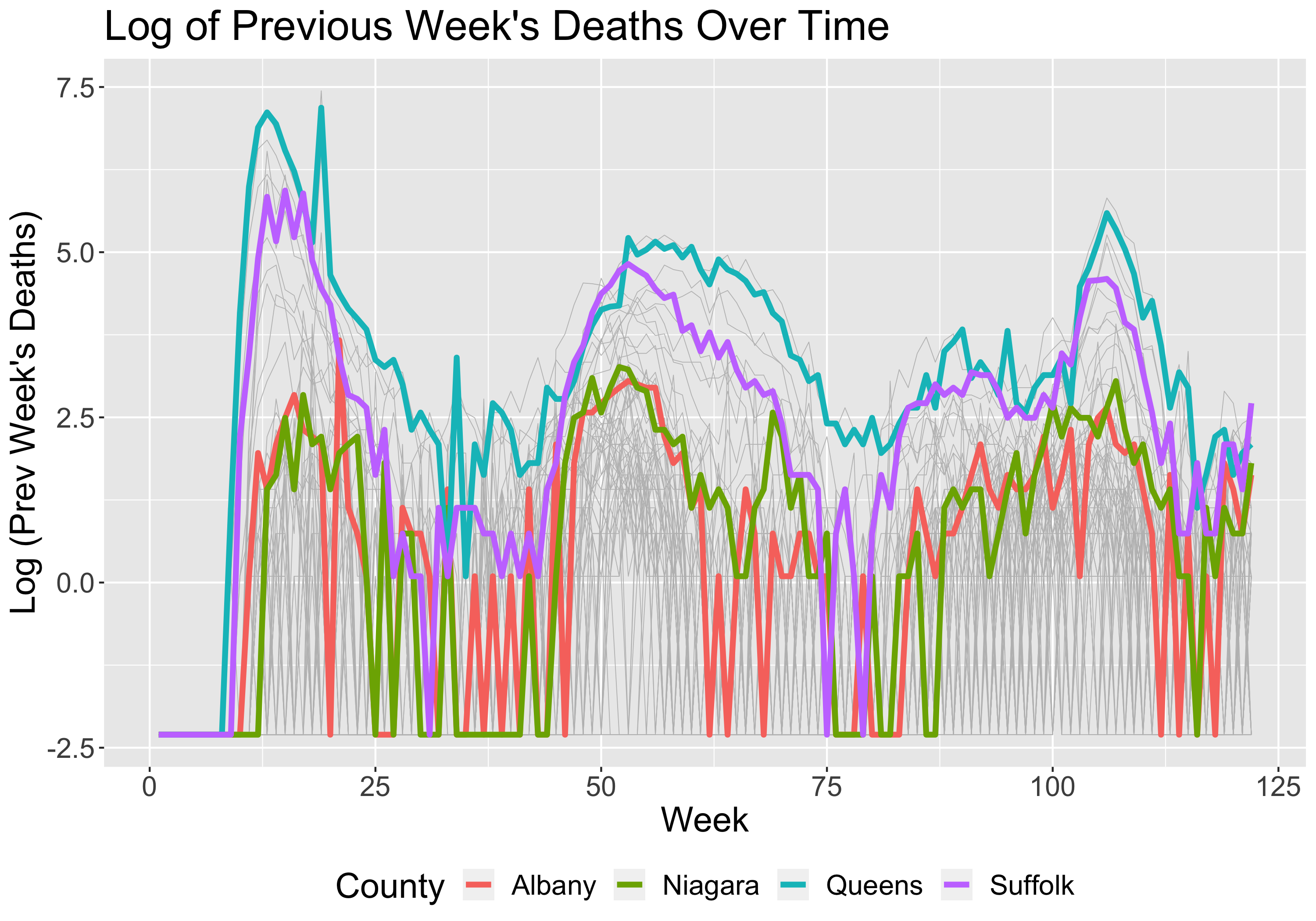}\\
(b)\\
\end{center}
\caption{Plots showing the values of vaccine prevalence (top) and log of the previous week's new deaths (bottom) across time.  The different lines represent the different counties, with four counties highlighted in color.}\label{fig:covariates}
\end{figure}

To understand how the covariate values change over time, we look at \Cref{fig:covariates}. Again, each line represents a unique county, with four illustrative counties highlighted in color. Like the transmission levels, the covariates display generally similar patterns in all counties, but there is quite a bit of variation in the values. The trend that the covariate values follow across time is different from the trend of the transmission levels; for example, vaccinations are non-decreasing, but the transmission levels both increase and decrease across the time period.  For this reason, a model that accounts for a changing relationship with the covariates across time is imperative. Because of the introduction of changepoints in the model, our methodology can account for this changing relationship in an apt manner. 

It is worth mentioning that we experimented with linear and quadratic trend functions as well, but since COVID-19 data shows ``wave'' like patterns, they were not found to be suitable for the study. Furthermore, the prevalence of the second dose of the COVID-19 vaccine was explored as an additional covariate. It was found to have a similar relationship as the first dose, thereby deemed not to be required in the model.

In the USA, responses to and policies for the COVID-19 pandemic varied widely from state to state, county to county, and even city to city. Thus, while the goal of our analysis was to examine all of New York state transmission levels together, we first explored whether analyses on individual counties would be more appropriate. In the interest of space, our analysis on this note is deferred to \Cref{appendix-exp}. Interestingly, we found that the data for individual counties was insufficient for estimation. For example, we found, first, that models for some counties (approximately half) did not converge even for a very large number of posterior samples. Additionally, the lack of data for a single county meant that we could also only fit these models to estimate a single changepoint, as there would not be enough information to estimate many (if any) changepoints beyond one.  Third, we found that for counties that did converge, estimates for changepoints and coefficients had very large uncertainty ranges and large overlap with other counties.  For these reasons, a spatio-temporal model that allows for borrowing information from neighbors is absolutely necessary.

\section{Methodology}
\label{sec:methods}

\subsection{Model development}\label{subsec:model}

Consider a categorical response variable $y(s_i,t)$ for locations $s_i \in \S \subset \R^2, \; i=1,\hdots,n$ and for time-points $t \in \Gamma$. While we primarily work with the regularly spaced set $\Gamma = \{1,\hdots,T\}$, we emphasize that the methodology would work for irregular spaced data as well. Likewise, although the COVID-19 data is not point-level, we find that the methodology works well for aggregate areal data, following similar works by other researchers \citep{jin2005generalized, bradley2015multivariate}.

For the response variable, let the total number of possible categories be represented by $m$. For convenience, we denote these categories as $1,2,\hdots,m$. Considering that the response is an ordered variable, we adopt the structure of ordinal probit models. See \cite{daykin2002analyzing} for a review of such models in the standard setting. To define the model in our case, let $\pi(s_i, t)$ be a latent random variable such that for $j=1,2,\hdots,m$,
\begin{equation}
\label{eq:latent-variable-ord}
    y(s_i,t) = j \; \text{ if }  \delta_{j-1} < \pi(s_i,t) \leqslant \delta_{j},
\end{equation}
where $-\infty=\delta_0<\delta_1=0<\delta_2<\hdots<\delta_{m-1}<\delta_{m}=\infty$ are the cut points determining the distribution of the categorical response.

Next, letting $t_0$ be the changepoint, the model for the latent variable is defined as
\begin{equation}
\label{eq:main-model2-ord}
    \pi(s_i, t) = \begin{cases} \mu(s_i,t) + u(s_i,t)  + \epsilon(s_i,t), & \text{for } t \leqslant t_0, \\ 
    \mu^*(s_i,t) + u(s_i,t) + v(s_i,t) + \epsilon^*(s_i,t), & \text{for }  t > t_0.
    \end{cases}
\end{equation}

In this model, $\mu(\cdot)$ and $\mu^*(\cdot)$ constitute the mean structure before and after the changepoint, respectively. The covariance structure in each of the two periods is written as a sum of the spatiotemporally dependent process(es) and a white noise process. The term $u(s_i,t)$ denotes the spatio-temporal effect before the changepoint, while $v(s_i,t)$ is the deviation in that effect following the structural break. Both processes are assumed to be Gaussian in nature. Finally, the white noise processes of the latent variables are defined by $\epsilon(s_i,t)$ and $\epsilon^*(s_i,t)$, and we assume them to be i.i.d. zero-mean Gaussian random variables with standard deviations $\sigma_{\epsilon}$ and $\sigma_{\epsilon^*}$, respectively. Hereafter, we shall use the notation $N_r(\cdot,\cdot)$ to denote $r$-variate Gaussian distributions.

In the model, both $\mu(\cdot)$ and $\mu^*(\cdot)$ are considered to be linear functions of the covariates, the effects of some of which can be spatially varying. To describe it mathematically, let $x_g(s_i,t)$ be the $g^{th}$ covariate ($1 \leqslant g \leqslant G$) whose effect is fixed for all locations, that is, the effect is not spatially varying. Similarly, for $1\leqslant h \leqslant H$, let $x_{g+h}(s_i,t)$ correspond to a covariate whose effect is considered to be spatially varying. Then, we write the mean functions as 
\begin{equation}
\label{eq:mean-struc}
\begin{split}
    \mu(s_i,t) &= \beta_0 + \sum_{g=1}^G \beta_g x_g(s_i,t) +   \sum_{h=1}^H \gamma_h(s_i)\,  x_{g+h}(s_i, t),\quad \text{for }\ t \leqslant t_0, \\
    \mu^*(s_i,t) &= \beta^*_0 + \sum_{g=1}^G \beta^*_g x_g(s_i,t) + \sum_{h=1}^H \gamma_h^{*'}(s_i)\, x_{g+h}(s_i, t),\quad \text{for } t> t_0;
\end{split}
\end{equation}
where $\beta_g$ and $\beta_g^*$ represent the coefficients for the fixed effect covariate before and after the change point, and $\gamma_h(s_i)$ and $\gamma_h^*(s_i)$ are location-specific coefficients in the two segments. In other words, $\bm \gamma_h = (\gamma_h(s_1), \dots, \gamma_h(s_n))'$ represents the coefficient vector accounting for spatial variation in the effect of the corresponding regressor. The vector $\bm\gamma_h^*$ has an identical interpretation as well. We assume these effects to be spatially correlated, following exponentially decaying covariance structure in a Gaussian setting. Specifically, for estimating the effect sizes in the Bayesian algorithm (to be discussed in detail in \Cref{subsec:bayesian}), we consider that $\bm \gamma_h \sim N_n(0,k\Omega^h_{s})$, and $\bm \gamma^{*}_h \sim N_n(0,k\Omega^{*h}_{s})$. It is worth mentioning that the vectors $\bm{\gamma}_1,\bm{\gamma}_2,\hdots,\bm{\gamma}_H$ are taken to be independent of one another. Now, the dispersion matrices $\Omega^h_{s}$ and $\Omega_{s}^{*h}$ are defined as
\begin{equation}
\label{eq:SIGMA-s}
    (\Omega^h_{s})_{ij}=\exp\{-\omega^h_{s}||s_i-s_j||\}, \; (\Omega^{*h}_{s})_{ij}=\exp\{-\omega^{*h}_{s}||s_i-s_j||\};
\end{equation}
where $\omega^h_{s}$ and $\omega^{*h}_{s} $ are the decay parameters before and after the changepoint, respectively. Note that this is a special case of Mat\'ern class of covariance functions, where the parameters reflect the extent of the decay in the correlation. For computation, the term $||s_i-s_j||$ is calculated by the great circle distance between two locations $s_i$ and $s_j$. 

For the spatio-temporal processes, akin to other studies in related applications  \citep[see, e.g.,][]{rawat2021spatio}, we assume separability for the correlation functions. Then, the spatial correlation function and the temporal correlation function are both considered to follow exponentially decaying structures. For the spatial correlation, we again use the great circle distance, whereas in the temporal correlation, the time difference $|t_1-t_2|$ is calculated by the lag between the time-points.  In other words, with decay parameters $\phi_{us}$, $\phi_{ut}$, $\phi_{vs}$, $\phi_{vt}$, one may write
\begin{equation}
    \label{eq:separable-cov}
    \begin{split}
        \corr\left(u(s_i,t_1),u(s_j,t_2)\right) &= \exp\left\{-\phi_{us}\norm{s_i-s_j}\right\} \times \exp\left\{-\phi_{ut}\abs{t_1-t_2}\right\}, \\
        \corr\left(v(s_i,t_1),v(s_j,t_2)\right) &= \exp\left\{-\phi_{vs}\norm{s_i-s_j}\right\} \times \exp\left\{-\phi_{vt}\abs{t_1-t_2}\right\}.
    \end{split}
\end{equation}

Throughout this article, we use $M=nT$ for the total number of observations. Let $\bm{Y}$ be the $M$-dimensional vector of $y(s_i,t)$ observations, first arranged by time-point and then by the index of the location. The vectors $\bm{\pi}$, $\bm{U}$, $\bm{V}$, $\bm{\epsilon}$, $\bm{\epsilon}^*$ are also created in the same fashion. The design matrix of regressor values is expressed as $X$. The aforementioned model using vector-matrix notation is then written as follows:
\begin{equation}
\label{eq:main-matmodel-ord}
    \bm{\pi} = \begin{cases} X \bm{\theta} + \bm{U}  + \bm{\epsilon}, & \text{for } t \leqslant t_0, \\ 
    X \bm{\theta}^* + \bm{U} + \bm{V} + \bm{\epsilon}^*, & \text{for } t > t_0, \end{cases}
\end{equation}
where $\bm \theta$ is the $k$-dimensional coefficient vector obtained by combining $(\beta_0,\beta_1,\hdots,\beta_G)'$ and the parameter vectors $(\bm{\gamma}_1',\bm{\gamma}_2',\hdots,\bm{\gamma}_H')'$. $\bm{\theta}^*$ is similarly defined.

Furthermore, from the above discussions, one may write $\bm{U} \sim N_M(0, \sigma^2_{u} \Sigma_{ut} \otimes  \Sigma_{us}))$, such that $(\Sigma_{ut})_{ij}=\exp\{-\phi_{ut}|t_i-t_j|\}$ and $(\Sigma_{us})_{ij}=\exp\{-\phi_{vs}||s_i-s_j||\}$. Similarly, we can define covariance matrices $\Sigma_{vt}$, $\Sigma_{vs}$ and write $\bm{V} \sim N_M(0, \sigma^2_{v} (\Sigma_{vt} \otimes  \Sigma_{vs}))$. For the white noise vectors, we know that $\bm{\epsilon} \sim N_M(0, \sigma^2_{\epsilon} \I )$ and $\bm{\epsilon}^* \sim N_M(0, \sigma^2_{\epsilon^*} \I)$, where $\I$ stands for an identity matrix of appropriate order.

\subsection{Bayesian estimation}
\label{subsec:bayesian}

We use a complete Bayesian framework to estimate the model parameters. To do that, suitable prior specifications are necessary. It should be noted that to avoid identifiability issues in the computations, following \cite{higgs2010clipped}, we set the variance parameters $\sigma^2_{u}$, $\sigma^2_{v}$, $\sigma^2_{\epsilon}$, and $\sigma^2_{\epsilon^*}$ equal to 1. We adopt the principles of Gibbs sampling for the posterior computation of the other parameters. Recall that it is a type of Markov chain Monte Carlo (MCMC) technique where each parameter is updated iteratively, following the full conditional posterior distribution given the values of the other parameters from the previous iteration. Interested readers may refer to \cite{gelfand2000gibbs} for relevant details on the origin and implementation of Gibbs sampling. To set up the steps of the sampler, we use the symbol $\mathcal{F}$ to indicate the information contained by all the parameters from the previous iteration except the parameter for which the full conditional posterior is calculated.

First, consider the latent vector $\bm\pi$ and let $\pi_{i}$ denote the $i^{th}$ element of it, while $\bm\pi_{-i}$ represents the vector $\bm\pi$ with all elements but the $i^{th}$ one. In the same spirit, $u_i$ and $v_i$ are defined for vectors $\bm U$ and $\bm V$ whereas $x_i'$ represents the $i^{th}$ row of the matrix $X$ defined in \cref{eq:main-matmodel-ord}. Let $t(i)$ be the time point corresponding to the same observation. Then, the full conditional distribution for $\pi_i$ is found by marginalizing over $\epsilon(s_i, t(i))$ and $\epsilon^*(s_i, t(i))$. Following the earlier notations, if $\mathcal{F}$ stands for $(\bm{Y},\bm{U},\bm{V},\bm{\theta},\bm{\theta}^*,\bm{\delta},t_0,\sigma_{\epsilon}^2,\sigma_{\epsilon^*}^2,\bm{\pi}_{-i})$, then it can be shown that
\begin{equation}
\label{eq:sigma_pi_ij-posterior-dist-ord}
\pi_{i}|\mathcal{F} \sim 
    \begin{cases} 
    TN(x'_i\bm{\theta} + u_i ,\sigma_{\epsilon}^2;\delta_{j-1},\delta_{j}), & \text{if } t(i) \leqslant t_0, \; y_i=j, \\
    TN(x'_i\bm{\theta}^*+u_i + v_{i},\sigma_{\epsilon^*}^2;\delta_{j-1},\delta_{j}), & \text{if } t(i) > t_0, \; y_i=j,
    \end{cases}
\end{equation}
where $TN(a,b;c,d)$ denotes the univariate truncated normal distribution, obtained by truncating a normal distribution with mean $a$ and variance $b$ in the interval $(c,d)$.

For each component in $\bm{\theta}$ and $\bm{\theta}^*$, we assume the uniform flat infinite-support prior distribution. Thus, posteriors for these parameters are dominated by the data, and no information is assumed in the prior distribution. The full conditional posterior distribution for these parameter vectors can then be written as
\begin{equation}
\label{eq:beta-posterior-ord}
\begin{split}
    \bm{\theta}| \mathcal{F} &\sim N_k\left( \Sigma_{\bm{\theta}}  \left[ \frac{X^{-'}(\bm{\pi}^{-}-\bm{U}^- )}{\sigma_{\epsilon}^2}   \right], \Sigma_{\bm{\theta}}  \right), \\
     \bm{\theta}^*| \mathcal{F} &\sim N_k\left( \Sigma_{\bm{\theta}^*}   \left[ \frac{X^{+'}(\bm{\pi}^{+}-\bm{U}^+ -\bm{V}^+)}{\sigma_{\epsilon}^2}   \right], \Sigma_{\bm{\theta}^*}  \right).
\end{split}
\end{equation}

In this equation, $\bm\pi^-$ and $\bm\pi^+$ represent the $\bm\pi$ vector before and after the changepoint, respectively. $\bm{U}^-$, $\bm{U}^+$, $\bm{V}^-$, $\bm{V}^+$, $\bm{Y}^-$, $\bm{Y}^+$, $T^-$, $T^+$, $X^-$, and $X^+$ are similarly defined. The dispersion matrices $\Sigma_{\bm{\theta}}$ and $\Sigma_{\bm{\theta}^*}$ are defined as
\begin{equation}
\label{eq:Sigma_beta-mat}
        \Sigma_{\bm{\theta}} =\left[ \frac{X^{-'}X^-}{\sigma_{\epsilon}^2}+ \frac{\Psi^{-1}}{k}\right]^{-1}, \;
        \Sigma_{\bm{\theta}^*} =\left[ \frac{X^{+'}X^+}{\sigma_{\epsilon^*}^2} + \frac{\Psi^{*-1}}{k}\right]^{-1}, 
\end{equation}
where $\Psi$ and $\Psi^*$ are block diagonal matrices with entries in the order $(\I_G,\Omega_{s_1},\Omega_{s_2},\hdots,\Omega_{s_H})$ and $(\I_G,\Omega^*_{s_1},\Omega^*_{s_2},\hdots,\Omega^*_{s_H})$, respectively. It is worth noting that the above conditional posteriors would be Gaussian (with different parameter values) even if one considers Gaussian priors for $\bm{\theta}$ and $\bm{\theta}^*$, instead of the aforementioned flat distributions.

Next, for calculating the posterior distributions of the cut points $\delta_2,\delta_3,\hdots,\delta_{m-1}$ we use the single variable slice sampling algorithm as used for other parameterizations of the cut points by \cite{heiner2022}. The R package ``diversitree" developed by \cite{FitzJohn2012} is used to implement the algorithm. A brief description of the single variable slice sampling algorithm is warranted here. The algorithm follows three main steps.  First, a random number $z$ is drawn from the vertical slice of $(0,g(r_0))$, where $r_0$ is the starting point for the random variable $R$ and $g(r_0)$ is its probability density function. Second, an interval of width $w$ is randomly positioned around $r_0$ and expanded until both ends are outside the horizontal slice where all values are equal to $z$. Third, a new point $r_1$ is randomly sampled from the interval obtained in the previous step of width $w$. For further details, please refer to \cite{neal2003slice}. 

For the prior distribution of $\delta_j$, the support must be between $\delta_{j-1}$ and $\delta_{j+1}$ since we are working with ordinal categorical data as discussed in \Cref{subsec:model}. We  use the uniform distribution between those two bounds as the prior distribution for $\delta_{j}$. Now, let $\bm{\pi}_j^-$ and $\bm{\pi}_{j+1}^-$ be the vectors that represent only those values from the vector $\bm{\pi}^-$ where the corresponding categories for the vector $\bm{Y}^-$ are $j$ and $j+1$, respectively. Similarly, $X_j^-$, $\bm{U}_j^-$, $X_{j+1}^-$, and $\bm{U}_{j+1}^-$ are defined. In an identical manner, let $\bm{\pi}_j^+$ and $\bm{\pi}_{j+1}^+$ be the vectors representing only those values from $\bm{\pi}^+$ where the corresponding categories for the vector $\bm{Y}^+$ are $j$ and $j+1$, respectively. Similarly, $X_j^+$, $\bm{U}_j^+$, $\bm{V}_j^+$, $X_{j+1}^+$, $\bm{U}_{j+1}^+$, and $\bm{V}_{j+1}^+$ are defined. Then, the full conditional posterior for $\delta_j$ can be written as
\begin{equation}
\label{eq:delta-post3}
\small
\begin{split}
    f(\delta_j|\mathcal{F}) & \propto I(\delta_{j-1}<\delta_j<\delta_{j+1})\intop_{\delta_{j-1}}^{\delta_{j}}  \exp{\bigg\{ \frac{-1}{2\sigma_{\epsilon}^2} \norm{\bm{\pi}_j^- -X_j^-\bm{\theta} - \bm{U}_j^- }^2 \bigg \} } {\mathrm{d} \bm{\pi}_j^-} \\
    &  \intop_{\delta_{j}}^{\delta_{j+1}}  \exp{\bigg\{ \frac{-1}{2\sigma_{\epsilon}^2} \norm{\bm{\pi}_{j+1}^- -X_{j+1}^-\bm{\theta} - \bm{U}_{j+1}^- }^2 \bigg \} } {\mathrm{d} \bm{\pi}_{j+1}^-} \\
    &\intop_{\delta_{j-1}}^{\delta_{j}} 
     \exp{\bigg\{ \frac{-1}{2\sigma_{\epsilon^*}^2}\norm{\bm{\pi}_j^+-X_j^+\bm{\theta}^* - \bm{U}_j^+- \bm{V}_j^+}^2 \bigg \} } {\mathrm{d} \bm{\pi}_j^+}\\
    &\intop_{\delta_{j}}^{\delta_{j+1}} 
     \exp{\bigg\{ \frac{-1}{2\sigma_{\epsilon^*}^2}\norm{\bm{\pi}_{j+1}^+-X_{j+1}^+\bm{\theta}^* - \bm{U}_{j+1}^+- \bm{V}_{j+1}^+}^2 \bigg \} } {\mathrm{d} \bm{\pi}_{j+1}^+}.
\end{split}
\end{equation}

It is easy to observe that the posterior distribution of $\delta_{j}$ does not have a closed form. To implement the slice sampling procedure, we simplify the above integral by making appropriate substitutions. It is worth mentioning that the multivariate slice sampling algorithm was also explored in this context, but it was computationally more expensive and less efficient for our algorithm compared to the single variable slice sampling procedure. 

We turn our attention to the space-time process vectors $\bm{U}$ and $\bm{V}$. Recall that the $M\times 1$ vector $\bm{Y}$ denotes the entire data and that $\bm{U}$ and $\bm{V}$ are the concatenated forms (by column) of the matrices
\begin{equation}
\label{eq:uv-structure}
    U_{T\times n} = \begin{pmatrix}
                    u_{s_1,1} & u_{s_1,2} & \hdots & u_{s_1,T}\\
                    u_{s_2,1} & u_{s_2,2} & \hdots & u_{s_2,T}\\
                    \vdots & \vdots & \vdots & \vdots\\
                    u_{s_n,1} & u_{s_n,2} & \hdots & u_{s_n,T}
                    \end{pmatrix}, 
    V_{T\times n} = \begin{pmatrix}
                    v_{s_1,1} & v_{s_1,2} & \hdots & v_{s_1,T}\\
                    v_{s_2,1} & v_{s_2,2} & \hdots & v_{s_2,T}\\
                    \vdots & \vdots & \vdots & \vdots\\
                    v_{s_n,1} & v_{s_n,2} & \hdots & v_{s_n,T}
                    \end{pmatrix}.
\end{equation}

It is critical to point out that the conventional posterior distributions for the entire vectors $\bm{U}$ and $\bm{V}$ would entail the inversion of $M\times M$ dimensional matrices in every iteration for every Markov chain during the Gibbs sampling procedure. This is computationally demanding, and to circumvent the problem, we follow a less expensive procedure by taking advantage of the separability of the spatio-temporal structure. Divide $\bm{V}$ into $\bm{V}_1,\bm{V}_2,\hdots,\bm{V}_T$ with $\bm{V}_t$ being the $t^{th}$  column of the $V$ matrix in \cref{eq:uv-structure}, and let $\bm{V}_{-t}=(\bm{V}_2,\hdots\bm{V}_{t-1},\bm{V}_1,\bm{V}_{t+1},\hdots,\bm{V}_T)$. Consider the partitioned structure 
\begin{equation*}
    \Sigma_{vt}=\begin{bmatrix}
             \Sigma_{11} & \Sigma_{12}\\
             \Sigma_{21} & \Sigma_{22}
             \end{bmatrix}
\end{equation*}
for the temporal correlation matrix, with subscript 1 denoting the correlation part corresponding to the $t^{th}$ time point and subscript 2 denoting the same for the rest of the time points.   Then, letting $\bm\mu_{vc} = (\Sigma_{12} \otimes \Sigma_{vs}) (\Sigma_{22}^{-1} \otimes \Sigma_{vs}^{-1})\bm{V}_{-t}$ and $\Sigma_{vc}=(\Sigma_{11} -\Sigma_{12}\Sigma_{22}^{-1} \Sigma_{21})\otimes \Sigma_{vs}$, multivariate normal distribution theory indicates that
\begin{equation}\label{eq:V1_conddist}
    \bm{V}_t|\bm{V}_{-t},\sigma_v^2 \sim N_n (\bm\mu_{vc},\sigma_v^2\Sigma_{vc}).
\end{equation}

Following the same idea for $\bm{U}$ as well, the full conditional posterior distributions are
\begin{equation}
\label{eq:v_i-posterior-ord}
\bm{V}_{t}| \mathcal{F} \sim
 \begin{cases}
    N_n (\mu_{vc},\sigma_{v}^2 \Sigma_{vc}
    ), & \text{for } t \leqslant t_0,\\
    N_n \biggl( \Sigma_{V_t}\left(\frac{\Sigma_{vc}^{-1}\bm\mu_{vc}}{\sigma_{v}^2} + \frac{\bm{\pi}_{t} - X_{t}\bm{\theta}^* - \bm{U}_t}{\sigma_{\epsilon^*}^2}\right),
    \Sigma_{V_t}\biggr), & \text{for } t > t_0,
    \end{cases}
\end{equation}

\begin{equation}
\label{eq:u_i-posterior-ord}
\bm{U}_{t}| \mathcal{F} \sim
 \begin{cases}
    N_n \biggl( \Sigma_{U_t}\left(\frac{\Sigma_{uc}^{-1}\bm\mu_{uc}}{\sigma_{u}^2} + \frac{\bm{\pi}_{t} - X_{t}\bm{\theta}}{\sigma_{\epsilon}^2}\right),
           \Sigma_{U_t}\biggr), & \text{for } t \leqslant t_0,\\
    N_n \biggl( \Sigma_{U_t}\left(\frac{\Sigma_{uc}^{-1}\bm\mu_{uc}}{\sigma_{u}^2} + \frac{\bm{\pi}_{t} - X_{t}\bm{\theta}^* - \bm{V}_t}{\sigma_{\epsilon}^2}\right),
           \Sigma_{U_t}\biggr), & \text{for } t > t_0,
    \end{cases}
\end{equation}
where $\Sigma_{V_t}$ and $\Sigma_{U_t}$ are defined to be
\begin{equation}
\label{eq:Sigma-V-U_i}
    \Sigma_{V_t} = \left[\frac{\Sigma_{vc}^{-1}}{\sigma_{v}^2}+\frac{\I_n}{\sigma_{\epsilon^*}^2}\right]^{-1}, \;
    \Sigma_{U_t} = \left[\frac{\Sigma_{uc}^{-1}}{\sigma_{u}^2}+\frac{\I_n}{\sigma_{\epsilon^*}^2}\right]^{-1}.
\end{equation}

Next, we look at the four decay parameters in the spatio-temporal covariance functions. One can calculate the posterior distributions of $\phi_{vt}$, $\phi_{vs}$, $\phi_{ut}$, and $\phi_{us}$, and show that they do not adhere to any known probability distribution. Thus, we use the single variable slice sampling algorithm as applied for $\delta_j$ before. It is particularly of note here that we estimate the decay parameters on a continuous scale, which is in stark contrast to the conventional cross-validation approach used by many other researchers. The problem with using a cross-validation method is that if there are more than two decay parameters being estimated, then only a small set of possible values for the parameters can be considered to handle the computational burden. For example, we look at the work of \cite{sahu2006spatio}, who used two spatio-temporal processes in their model for analyzing air pollution levels. However, in their study, they had to find the optimal values of the decay parameters from a set of only five values for the spatial decay parameters and only three values for the temporal decay parameters. Clearly, this limits the algorithm's ability to explore all possible values for the decay parameters and runs  the risk of not exploring the actual sample space (a four-dimensional cube for our model). We take advantage of the aforementioned slice sampling procedure to alleviate this computational burden in estimating the decay parameters. For the priors, we choose uniform distributions between 0 and 3 to obtain the full conditional posterior distributions as follows
\begin{equation}
\label{eq:phi-posterior-ord}
    \begin{split}
        \phi_{vt}|\mathcal{F} &\propto |\Sigma_{vt}|^{-n/2} \exp \Bigg\{\frac{- \bm{V}'(\Sigma_{vt}^{-1} \otimes \Sigma_{vs}^{-1}) \bm{V}}{2\sigma_{v}^2}\Bigg\} I(0<\phi_{vt}<3),\\
        \phi_{vs}|\mathcal{F} &\propto |\Sigma_{vs}|^{-T/2} \exp \Bigg\{\frac{- \bm{V}'(\Sigma_{vt}^{-1} \otimes \Sigma_{vs}^{-1}) \bm{V}}{2\sigma_{v}^2}\Bigg\} I(0<\phi_{vs}<3),\\
        \phi_{ut}|\mathcal{F} &\propto |\Sigma_{ut}|^{-n/2} \exp \Bigg\{\frac{- \bm{U}'(\Sigma_{ut}^{-1} \otimes \Sigma_{us}^{-1}) \bm{U}}{2\sigma_{u}^2}\Bigg\} I(0<\phi_{ut}<3),\\
        \phi_{us}|\mathcal{F} &\propto |\Sigma_{us}|^{-T/2} \exp \Bigg\{\frac{- \bm{U}'(\Sigma_{ut}^{-1} \otimes \Sigma_{us}^{-1}) \bm{U}}{2\sigma_{u}^2}\Bigg\} I(0<\phi_{us}<3).
    \end{split}
\end{equation}

In an exactly similar fashion, we can also show that for the decay parameters of the spatially-varying coefficients,
\begin{equation}
\label{eq:omega-posterior-ord}
    \begin{split}
     \omega^h_{s}|\mathcal{F} &\propto |\Omega^h_{s}|^{-1/2} \exp \Bigg\{\frac{- \bm{\gamma_h}'(\Omega_{s}^{h})^{-1} \bm{\gamma_h}}{2k}\Bigg\} I(0<\omega^h_{s}<3),\\
     \omega^{*h}_{s}|\mathcal{F} &\propto |\Omega^{*h}_{s}|^{-1/2} \exp \Bigg\{\frac{- \bm{\gamma_h}^{*'}(\Omega_{s}^{*h})^{-1}\bm{\gamma_h}^*}{2k}\Bigg\} I(0<\omega^{*h}_{s}<3).
    \end{split}
\end{equation}

Finally, for calculating the full conditional posterior distribution of the changepoint $t_0$, we consider the set $S_T=\{0, 1, 2, \hdots, T\}$ and use a discrete uniform prior distribution. Let $f_{S_T}(t_0)$ denote this prior. Note that the endpoints are included in the sample space so as to enable the algorithm to converge to any of the endpoints, which would effectively suggest no changepoint. Then, the Gibbs sampling step corresponding to the parameter $t_0$ requires the conditional posterior distribution, which is described by the probability mass function as
\begin{equation}
\label{eq:t_0-posterior-ord}
\begin{split}
    f(t_0|\mathcal{F}) \propto \; & I(t_0 \in S_T) (2\pi \sigma_{\epsilon}^2)^{\frac{-nT^-}{2}}\exp \bigg \{ \frac{-1}{2\sigma_{\epsilon}^2} \norm{\bm{\pi}  ^{-}-(X^-\bm{\theta}+\bm{U}^-)}^2 \bigg \} \\
    & (2\pi \sigma_{\epsilon^*}^2)^{\frac{-nT^+}{2}}\exp \bigg \{ \frac{-1}{2\sigma_{\epsilon^*}^2} \norm{\bm{\pi}^{+}-(X^+\bm{\theta}^*+\bm{U}^++\bm{V}^+)}^2 \bigg \}.
\end{split}
\end{equation}

\subsection{Multiple changepoint detection in COVID-19 Data}
\label{subsec:app}

In the previous two subsections, we discussed the Bayesian estimation of the proposed model of changepoint detection in a general sense. Here, we explain how the model is applied to the COVID-19 data described in \Cref{sec:data}. For this application, as mentioned already, we use the logarithmic transformation of the number of new deaths in the previous week as a non-spatially varying covariate. On the other hand, the prevalence of the first COVID-19 vaccine dose is used as a spatially-varying coefficient in the model. Let us use $\log d(s_i,t-1)$ to denote the first covariate at time $t$ and location $s_i$, whereas $\bm x(t)=(x(s_1,t),\hdots,x(s_n,t))'$  is the $n$-dimensional vector of the vaccination prevalence in the set of locations. Then, the mean structure of the latent variable can be written as
\begin{equation}
\label{eq:mean-struc-app}
\begin{split}
    \mu(s_i,t) &= \beta_0 + \beta_1 \log d(s_i,t-1) + \bm \gamma(s_i) x(s_i,t),\quad \text{for }\ t \leqslant t_0, \\
    \mu^*(s_i,t) &= \beta^*_0 + \beta^*_1 \log d(s_i,t-1) + \gamma^{*}(s_i) x(s_i,t),\quad \text{for } t> t_0.
\end{split}
\end{equation}

For the covariate related to the previous week's death, it is expected that it contributes positively to the spreading of the disease. It is connected to the idea that death by COVID-19 corresponds to a higher viral count, which increases the chance of spreading the disease to another human \citep{pujadas2020sars}. However, as more people are vaccinated, we expect this effect to be negligible, as there will be fewer deaths after vaccination. Regarding the second covariate, we expect a significantly negative coefficient in the first phase of the vaccination, which would provide evidence of the effectiveness of COVID-19 vaccines in decreasing the spread of the pandemic. As the vaccination picks up, herd immunity is likely to be acquired. For example, \cite{macintyre2022modelling} demonstrated in their research that Australia achieved herd immunity with the vaccination coverage of around 66\% population. Once that status is reached, the effect of vaccines should not be prominent in explaining the spread of the disease. Furthermore, taking inspiration from \cite{utazi2018high} who studied the spatially-varying impact of measles vaccination in different countries, we hypothesize that different counties in New York state may have experienced unequal effects of vaccines in its relationship with the spread of the disease. This is, in fact, the primary motivation behind using the spatially-varying coefficients for the first-dose vaccination. 

Another crucial aspect of our proposed methodology is the ability to assess whether the estimated changepoint is indeed significant. Following the convention in Bayesian literature, we make use of the Bayes factor. The Bayes factor allows us to compare a model that does not have any changepoint in the structure to the model with a changepoint. Let $\mathcal{M}_1$ denote our proposed model with a changepoint (see \cref{eq:latent-variable-ord} and \cref{eq:main-model2-ord}) and $\mathcal{M}_2$ be the other model obtained by fitting the model with no changepoint (i.e., the first line before the changepoint in \cref{eq:main-model2-ord} for the entire data). The likelihood of the data given a particular model is then
\begin{equation*}
\label{eq:likeli-Y-model}
\begin{split}
    f(\bm{Y}|\mathcal{M}_i) &= \int f(\bm{Y},\bm{\theta},\bm{\theta}^*,\bm{U},\bm{V}|\mathcal{M}_i) \mathrm{d} \bm{\theta} \mathrm{d} \bm{\theta}^*\mathrm{d} \bm{U}\mathrm{d} \bm{V}\\
    &= \int f(\bm{Y}|\bm{\theta},\bm{\theta}^*,\bm{U},\bm{V},\mathcal{M}_i)f(\bm{\theta},\bm{\theta}^*,\bm{U},\bm{V}|\mathcal{M}_i) \mathrm{d} \bm{\theta} \mathrm{d} \bm{\theta}^* \mathrm{d} \bm{U}\mathrm{d} \bm{V}\\
    &= \mathbb{E}_{\bm{\theta},\bm{\theta}^*,\bm{U},\bm{V}} \left[f(\bm{Y}|\bm{\theta},\bm{\theta}^*,\bm{U},\bm{V},\mathcal{M}_i)\right] \\
    &\approx \frac{1}{m}\sum_{j=1}^m f\left(\bm{Y}|\bm{\theta}^{(j)},\bm{\theta}^{*(j)},\bm{U}^{(j)},\bm{V}^{(j)},\mathcal{M}_i\right),
\end{split}
\end{equation*}
where $\bm{\theta}^{(j)} \sim f(\bm{\theta}|\mathcal{M}_i))$, $\bm{\theta}^{*(j)} \sim f(\bm{\theta}^{*}|\mathcal{M}_i))$, $\bm{U}^{*(j)} \sim f(\bm{U}|\mathcal{M}_i))$, $\bm{V}^{*(j)} \sim f(\bm{V}|\mathcal{M}_i))$ are realizations from the model-specific distributions. Interestingly, often the likelihood values are intractable if one uses non-informative priors, as in this case. As suggested by \cite{newton1994approximate} and \cite{risser2019nonstationary}, in such cases, one can leverage the posterior distributions of the parameters to find the likelihood with the help of the harmonic mean. Specifically, if $\bm{\beta}_{post}^{(j)}$ denotes the realization of the parameter vector from the $j^{th}$ posterior sample (similarly for other parameters), then the likelihood is obtained as 
\begin{equation}
    f(\bm{Y}|\mathcal{M}_i) \approx \Bigg[ \frac{1}{m}\sum_{j=1}^m \frac{1}{f(\bm{Y}|\bm{\beta}_{post}^{(j)},\bm{\beta}^{*(j)}_{post},\bm{U}_{post}^{(j)},\bm{V}_{post}^{(j)},\mathcal{M}_i)} \Bigg]^{-1}.
\end{equation}

Once the likelihood values are computed, we use the Bayes factor to decide if the estimated changepoint is significant in our problem. Following \cite{kass1995bayes}, we use the cutoff of 100 for the Bayes factor to deduce whether there is decisive evidence in favor of the changepoint model. Recall that the Bayes factor for the model $\mathcal{M}_1$ with respect to $\mathcal{M}_2$ is given by
\begin{equation}
\label{eq:bayes-factor}    
    BF_{12} = \frac{f(\bm{Y}|\mathcal{M}_1)}{f(\bm{Y}|\mathcal{M}_2)}.
\end{equation}

The above procedure is not only useful in finding whether there is a single significant changepoint, but it is also useful for finding if there are multiple changepoints in the ordered categorical spatio-temporal data. To that end, we propose a binary segmentation-type algorithm. This technique has been used in the literature to detect multiple changepoints in different problems \citep{fryzlewicz2014wild, cho2015multiple}. To implement the algorithm in our context, first, we run the aforementioned procedure to find a changepoint, if any, in the entire time period. In all instances of Gibbs samplers in this work, we monitor the convergence through the Gelman-Rubin diagnostic (\cite{gelman1992inference}), running three Markov chains simultaneously. Now, if the Bayes factor approach provides decisive evidence in favor of the changepoint, we divide the data into two segments with respect to the temporal domain. Then, the same algorithm is run on each of the two segments, and we proceed recursively until no significant changepoint is left in the data. From a pragmatic standpoint, as an additional stopping rule, we impose the restriction that any segment identified by the changepoint(s) must be at least three months (equivalently, 12 weeks) long. Thus, if there are no more than 24 weeks of data for a particular segment, we do not run the algorithm further. Also, if a changepoint is found to be too close to either of the endpoints, it is ignored based on the same restriction.

We present the pseudo-code of the entire procedure of multiple changepoint detection in ordered categorical spatio-temporal data in Algorithm \ref{alg:estim}. Note that the stopping criteria $\mathcal{C}$ implies that either the Bayes factor approach does not support the existence of a changepoint or that every segment has no more than the required amount of data. We emphasize that this is a subjective choice and can be easily adjusted under other considerations.

\begin{algorithm}
\caption{Multiple changepoint detection in spatio-temporal categorical data.}
\label{alg:estim}
    \SetKwInOut{Input}{Input}
    \SetKwInOut{Output}{Output}
    \SetKwInOut{Initialization}{Initialization}
    \Input{Data vector $\bm Y$, design matrix $X$, and stopping criteria $\mathcal{C}$.}
    \Output{Posterior distribution for all the parameters and changepoints.}
    \Initialization{Set $\bm{U}=0$ and $\bm{V}=0$, take random feasible values for $\phi_{us}$, $\phi_{ut}$, $\phi_{vs}$, $\phi_{vt}$,$\omega_{s}$, $\omega^{*}_{s}$, $\bm{\beta}$, $\bm{\beta}^*$, $\delta_2$, $\delta_3$, $t_0$ for three Markov chains.}
    \While {$\mathcal{C}$ not true}{
    (a) Implement model $\mathcal{M}_1$ (with changepoint) as described above, find the posterior samples, estimate the parameters and compute the estimated likelihood for the model. \;
    (b) Implement model $\mathcal{M}_2$ (without the changepoint) as described above, find the posterior samples, estimate the parameters and compute the estimated likelihood for the model.\;
    (c) Compute the Bayes factor of $\mathcal{M}_1$ with respect to $\mathcal{M}_2$. \;
    (d) Divide the dataset into two segments based on the estimated changepoint in $\mathcal{M}_1$. \;
    (e) \eIf{Bayes factor $>$ 100 \& sufficient data are available for one or both segments.}{Use the data from the segment(s) where enough data are available; and go to step (a).\;}{
    break\;
    }
    }
\end{algorithm}

\section{Results}
\label{sec:results}

As mentioned in \Cref{sec:data}, our objective is to utilize the proposed model to detect the changepoints in the spatio-temporal spread of COVID-19 by considering its ordered categorical nature. Recall that the data used in this study comprise 122 weeks of information (20$^{th}$ January 2020 to 16$^{th}$ May 2022) from 62 counties in the state of New York. In this section, we first discuss the changepoints identified by the algorithm and discuss their implications. We then provide an in-depth look at the relationship between the covariates and the spread of the pandemic. Finally, we examine the extent of spatial and temporal dependence in the propagation of the disease across different time periods. 

In this application, using the steps in Algorithm \ref{alg:estim}, the proposed methodology detects three changepoints. \Cref{tab:cp-est} shows the estimated changepoints with the corresponding Bayes factors in all stages of the analysis, including those that were not found to be significant according to the criteria. The first changepoint is estimated at the $57^{th}$ week, and the corresponding Bayes factor is above the cutoff, indicating evidence for a changepoint. In the second stage, the data set is divided into two time periods, and the proposed model is applied to each separate time period. For the first time period, i.e., for the $1^{st}$ to $57^{th}$ weeks of the data, because the vaccination started only at the end of the time period (approximately week 50), we had to remove it as a covariate. The Bayes factor here again supports the presence of changepoint, which is obtained at the $36^{th}$ week. Likewise, for the second time period, i.e.,\ the $58^{th}$ to $122^{nd}$ weeks, the $96^{th}$ week is found to be a significant changepoint. Subsequently, in the third stage of the binary segmentation algorithm, when we fit the model again on the data from $1^{st}$ to $36^{th}$ week, the detected changepoint is towards the beginning of the time period with a logarithmic Bayes factor of $-72$, suggesting no changepoint in the structure. A similar conclusion is reached when the model is run on the data from the $58^{th}$ to $96^{th}$ week. Meanwhile, in the time period from $97^{th}$ to $122^{nd}$ week, the changepoint is found to be too close to the beginning of the time period and is ignored based on our restriction. Finally, note that we do not run the algorithm for the remaining time period because there is insufficient data according to our assumption mentioned earlier. Thus, there are three changepoints in this data set, and they are on $21^{st}$ September 2020, $15^{th}$ February 2021, and $15^{th}$ November 2021.

\begin{table}[ht]
\caption{Changepoints and respective Bayes factors for the proposed algorithm. The first column indicates the stages of binary segmentation. Bolded changepoints indicate that it was identified as present/significant in the data.}
\label{tab:cp-est}
\centering
\begin{tabular}{llccc}
  \hline
  Stage & Time horizon & Changepoint & Date & Log Bayes Factor \\ 
  \hline
 1 & $1^{st}$ to $122^{nd}$ week  & \textbf{$\bm{57^{th}}$ week} & $15^{th}$ Feb 2021 & 44 \\ 
  \hline
  2 & $1^{st}$ to $57^{th}$ week & \textbf{$\bm{36^{th}}$ week}& $21^{st}$ Sep 2020  & 64 \\
    & $58^{th}$ to $122^{nd}$ week & \textbf{$\bm{96^{th}}$ week}& $15^{th}$ Nov 2021 & 1118 \\
  \hline
  3 & $1^{st}$ to $36^{th}$ week & $8^{th}$ week & $9^{th}$ Mar 2020 & $-72$ \\ 
  & $37^{th}$ to $57^{th}$ week & Insufficient data & & \\ 
  & $58^{th}$ to $96^{th}$ week & $79^{th}$ week& $19^{th}$ Jul 2021 & $-102$ \\
   & $97^{th}$ to $122^{nd}$ week & Not found & & \\ 
  \hline
\end{tabular}
\end{table}

\begin{figure}[!hbt]
\begin{center}
\includegraphics[width = 0.8\textwidth,keepaspectratio]{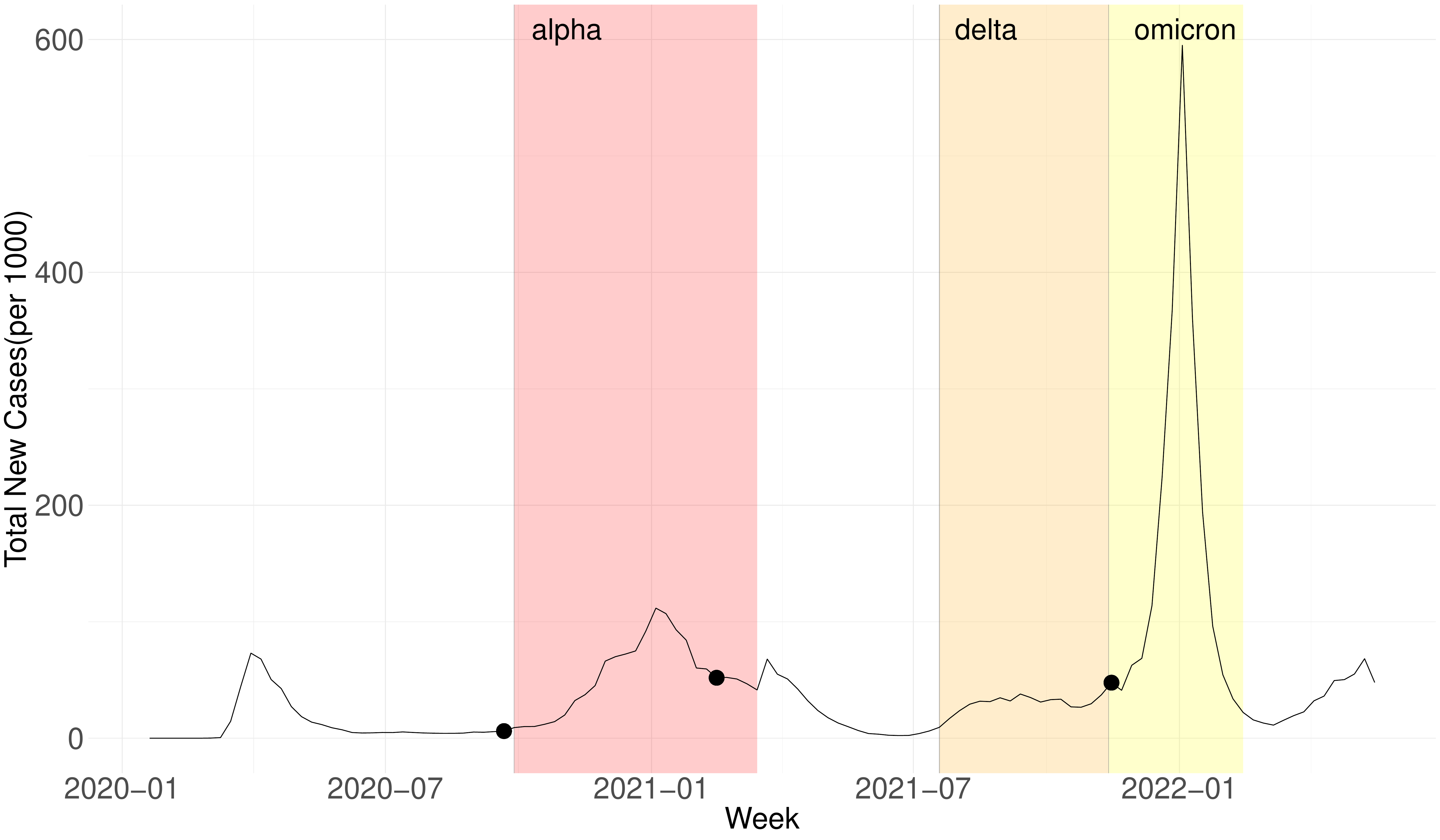}
\end{center}
\caption{COVID-19 waves corresponding to new cases for New York State. The graph's black dots denote the proposed model's estimated changepoint.}
\label{fig:NY-covid-wave}
\end{figure}

To put this into context, \Cref{fig:NY-covid-wave} shows the three changepoints (represented by black dots) with the total number of new cases in the state. This figure shows that the changepoint on $15^{th}$ February 2021 is towards the end of the winter alpha wave of COVID-19 in the USA. It is also the time when vaccinations picked up rapidly in the country, and around 11\% of the New York state population received the first dose.  In contrast, the changepoint on $21^{st}$ September 2020 coincides with the period when the same wave was about to start in the state.  It was undoubtedly one of the most challenging times for the state, and that is well reflected by the two changepoints. Subsequently, we notice that the period of the delta variant did not experience any significant change in the way the disease spread, but the beginning of the wave of the Omicron variant that peaked around the end of January aligns perfectly well with the third changepoint, $15^{th}$ November 2021. This clearly demonstrates strong support for the proposed model to correctly detect the different types of COVID-19 waves in New York.

Next, we discuss the effect sizes of the covariates, obtained from fitting the proposed model to the different time periods. We report the posterior means and the 95\% credible intervals in \Cref{table:est-cp}. The spatially-varying coefficients for vaccinations are shown in \Cref{fig:covariates} and discussed in a later paragraph. These values are obtained based on the time periods formed by the changepoints, as discussed before. Notably, for both the intercept and the coefficient for $\log d(s_i, t-1)$, there are quite different estimates in the four segments. The intercept for the first phase, when the pandemic was gradually picking up, is much lower. This makes sense, given the relatively lower number of reported COVID-19 cases in this period. Then, the alpha wave started running wild during the $37^{th}$ to $57^{th}$ week period. As vaccines did not provide much intervention until then, the intercept's estimated value is found to be relatively high. In the following segment, with more people getting vaccinated, the number of new cases decreased, thereby justifying a lower value for the estimated intercept. Contrary to that, in the last period, a generally higher average in the latent variable is observed. It can be substantiated by the rise in the new COVID-19 cases, with the omicron variant escaping the vaccination-induced immunity to take over the entire country. Interested readers may refer to \cite{tian2021global} and \cite{ao2022sars} for some related discussions in this regard.

\begin{table}[!ht]
\small
\centering
\caption{Parameter estimates and corresponding credible intervals when the proposed model is fitted in different segments.} 
\label{table:est-cp}
\begin{tabular}{lccc}
 \hline
Variable & Week Range & Posterior mean & $95\%$ credible interval  \\ 
  \hline
  Intercept  &  1-36   & $1.079$   & $(0.467, 1.615)$       \\
             &  37-57   & $7.713$   & $(5.483, 9.525)$       \\
   &  58-96  & $6.279$   & $(5.124,7.406)$       \\
    & 97-122    & $7.978$   & $(6.520,9.321)$       \\
 \hline
 log(Previous week's death) & 1-36 & $1.101$    & $(0.679, 1.497)$  \\
                & 37-57 & $0.296$    & $(0.071, 0.533)$  \\
& 58-96 & $0.273$    & $(0.157,0.389)$ \\
& 97-122     & $-0.146$    & $(-0.357,0.061)$ \\
 \hline
  Spatial decay parameter  &  1-57  & 0.0108  &(0.0081,0.0129)\\
    & 58-122    & 0.0021     & (0.0016,0.0026)   \\
  \hline
  Temporal decay parameter  &  1-57  & 0.264     & (0.206,0.315) \\
     & 58-122   & 0.203      & (0.165,0.245) \\
  \hline
  \end{tabular}
\end{table}

For the coefficient of $\log d(s_i,t-1)$, initially, when there was no available vaccine, the variable was significantly related to a rise in the new cases of COVID-19 infections. In fact, for the period of $1^{st}$ to $36^{th}$ week, the coefficient for the previous week's number of deaths is the highest among all periods, indicating a strong positive relationship to the transmission level of COVID-19. During this phase, the state of New York reported around 6000 deaths on a weekly basis. In the next segment, the state was experiencing the alpha wave. The number of deaths came down in this stage, which points to less severity of the disease, and naturally explains the lower value of the estimated coefficient. Furthermore, once a considerable proportion of the population was vaccinated in the third time period, we see the coefficient estimates for the death numbers going down significantly. This may be perceived as the effectiveness of the vaccines. Finally, for the period between $97^{th}$ and $122^{nd}$ week, the coefficient for the death numbers is found to be insignificant as the omicron variant was milder in terms of causing severe illness but was highly infectious (explaining the large intercept term in this segment). Thus, in addition to capturing changepoints, the proposed model has appropriately captured this effect as well. 

\begin{figure}[!hbt]
\begin{center}
\includegraphics[width = 0.8\textwidth,keepaspectratio]{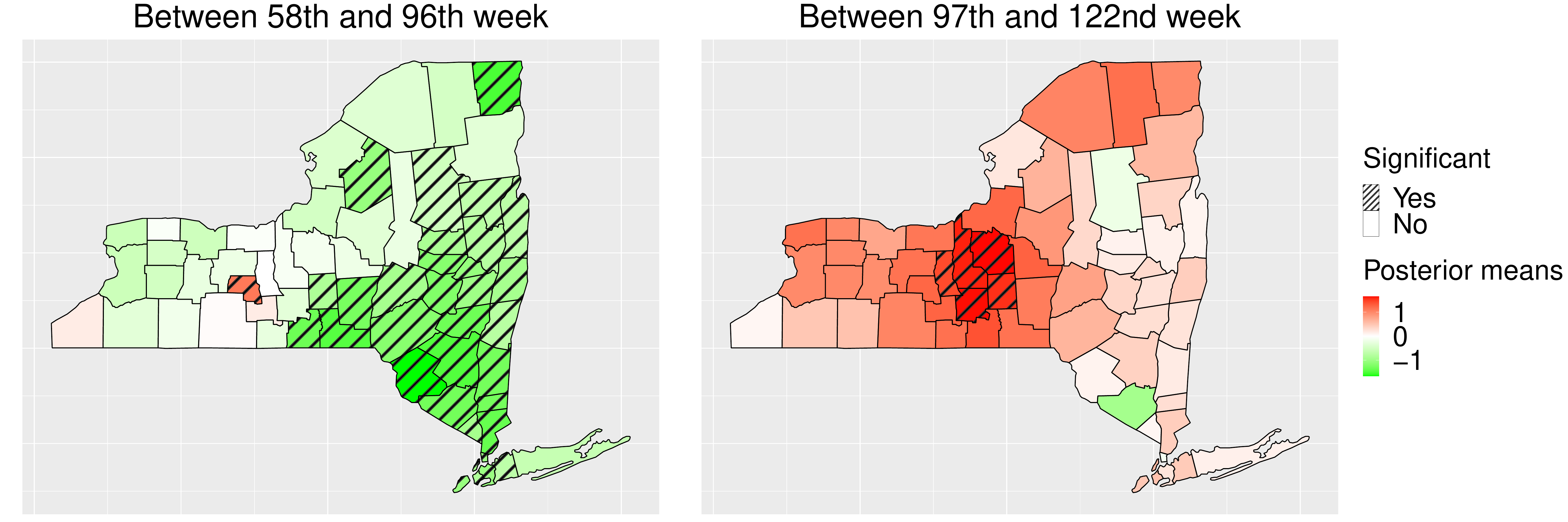}
\end{center}
\caption{Spatially varying coefficients (posterior means) for the vaccination prevalence in the last two segments obtained via the changepoints. Significance is equivalent to the credible interval being sufficiently away from zero.}
\label{fig:NY-sp-varying}
\end{figure}

Let us now focus on the effect of the prevalence of vaccination. It is known that the vaccine was offered only to hospital workers and older people in the initial days. In that light, relevant data are unavailable before the $57^{th}$ week, and this covariate has to be omitted from the model. Therefore, we present the spatially varying effect of the vaccination in the other two segments in \Cref{fig:NY-sp-varying}. The plots show the parameters' posterior means, and stripes indicate that the credible interval is sufficiently away from zero. 

When vaccination rates picked up in the period of $58^{th}$ to $96^{th}$ week, it was significantly related to a decrease in the transmission level of COVID-19 for all counties in New York. For 33 of the 62 counties, this coefficient is found to be significant. It includes almost all counties in the eastern part of the state, which are typically more populous than the rest. One can thus argue that the vaccination was more effective in the initial phase in more populated counties. In contrast, for the period of $97^{th}$ to $122^{nd}$ week, we see that vaccination is generally not significant. One possible justification is the aspect of herd immunity. By $22^{nd}$ November 2021, 75\% of the state population received their first dose, and by $16^{th}$ May 2022, it was more than 85\% of the population. As discussed in \Cref{subsec:app}, around two-thirds of population coverage is required to achieve herd immunity. According to \cite{who2021} too, 70\% of the population should be vaccinated to ensure herd immunity against COVID-19. Therefore, after achieving this much population coverage, the effect of the vaccination became irrelevant, until there was no resistance for the omicron variant \citep{ao2022sars}. 

\begin{figure}[!hbt]
\begin{center}
\includegraphics[width = 0.8\textwidth,keepaspectratio]{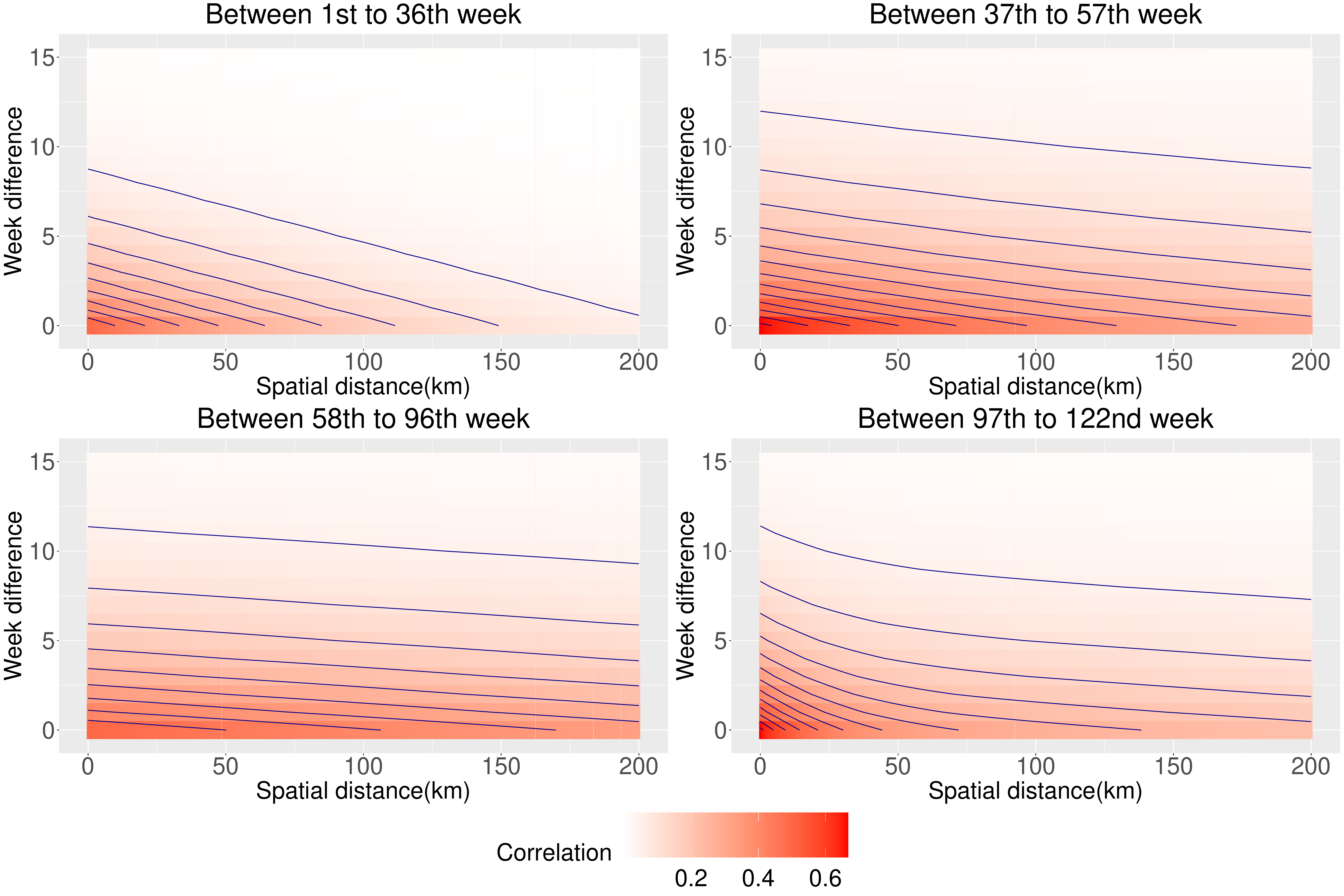}
\end{center}
\caption{Estimated correlation for the latent process $\pi(s_i,t)$ in four segments. The contour lines represent the points that have the same correlation values.}
\label{fig:corr-plot}
\end{figure}

As a last piece of the discussion, we look at the spatio-temporal dependence in the four segments created by the changepoints. To this end, heatmaps of the estimated spatio-temporal correlation for the latent process $\pi(s_i,t)$ for different periods are displayed in \Cref{fig:corr-plot}. The correlations are computed based on the posterior means of the spatial and temporal decay parameters and are presented as functions of distance and time-lag.

The correlation is estimated to be the lowest in the first time period compared to all other time periods. A probable reason is that lockdowns were enforced in New York very early when the COVID-19 prevalence was quite low. After that, restrictions were relaxed slowly in a four-phase technique \citep{huschblackwell}.  Thus, as the pandemic progressed and the prevalence kept increasing during the alpha wave ($37^{th}$ to $57^{th}$ weeks), we see that there is a strong spatio-temporal dependence. It was, in fact, the largest correlation among all time periods. Afterward, between $58^{th}$ and $96^{th}$ week, it can be seen from the plots that there is a drop in spatial and temporal dependence. This can be attributed to the fact that during this time period, vaccination rates started picking up everywhere, as many vaccines were made available to all adults of different age groups. Finally, we see a further drop in the spatial and temporal dependence pattern for the last time period. As pointed out before, the omicron variant prevailing in this phase was milder than the previous counterparts and did not cause severe illness. The population had also arguably reached herd immunity. It can be the reason behind the drop in spatial dependence, whereas the drop in the temporal correlation can be connected to the fact that illness due to the omicron variant persisted for a shorter duration than the delta variant, as explicated by \cite{menni2022symptom} and \cite{vitiello2022advances}.

\section{Conclusion}
\label{sec:conclusion}

In this work, we have developed a novel methodology to detect changepoints in an ordinal categorical spatio-temporal data set. We also provided complete estimation details and considerations for the various model parameters. The model is applied to COVID-19 data from the state of New York, by converting weekly new cases for every county into ordered categories (following CDC guidelines), signifying the transmission levels of COVID-19. Our model is defined through an appropriately structured latent process that can quantify the spatial and temporal dependence in the spread of the disease. It is implemented through a complete Bayesian framework. Our analysis identifies interesting changes in the spatial and temporal patterns of the pandemic, which are closely aligned with the infectivity of specific virus variants. Taking advantage of our proposed model, we also illustrate how vaccination is related to the transmission levels in different counties over the time period. 

One interesting finding is the evidence of varying spatial dependence in various time segments, which may be attributed to the movement of people and the vaccination rate. Due to the unavailability of movement data across different counties of New York state, this is beyond the scope of the paper. Nonetheless, this could be a valuable extension of our work to understand the infection dynamics better. In a similar vein, restrictions, and lockdowns imposed by the government are expected to impact the spread of COVID-19. Such covariates related to policy implementation have not been incorporated into our analysis, once again, because of the lack of relevant data at an appropriate granularity. Naturally, it poses another idea of an extension to the current approach. 

It is imperative to point out that not only the application of our methodology to COVID-19 data was useful for providing additional insights into a world-changing event, but the proposed model would also be helpful in other applications. For example, it can be utilized in epidemiological (e.g., other infectious disease transmission levels), environmental (e.g., precipitation levels or soil organic carbon concentration levels), education (e.g., school grades over time), social network (e.g., level of connectedness of individuals or companies) studies, and more.  

Finally, we note that the proposed methodology in this article has limitations in terms of computational burden in the case of a bigger data set, e.g., if one wants to look at all counties of the USA in a single model. Particularly, it would be an intriguing exercise to estimate the changepoints for different regions across the country and assess how similar or different the spread of COVID-19 was in different periods. Although our approach works in theory, in such cases, applying a model at this scale would require a significant approximation in the Bayesian algorithm.

\bibliography{references}

\newpage

\appendix
\numberwithin{equation}{section}
\numberwithin{table}{section}
\numberwithin{figure}{section}


\section{Exploratory Analysis for Individual Counties}
\label{appendix-exp}

In the United States, responses to and policies for the COVID-19 pandemic varied widely from state to state, county to county, and even city to city. Thus, while the goal of our analysis was to examine all of New York state transmission levels together, we first explored whether analyses on individual counties would be more appropriate.  

To approach this, we fit the proposed model to each county, removing spatial dependence (or, fixing $\Sigma_{us} = 1$).  Specifically, $\mathbf{U} \sim N(0, \sigma^2_U (\Sigma_{ut} \otimes \Sigma_{us}))$ was simplified to $\mathbf{U} \sim N(0, \sigma^2_U \Sigma_{ut})$, and similarly for $\mathbf{V}$.

We found, first, that models for some counties (approximately half) did not converge in the pre-defined time allotted.  For this reason, the spatio-temporal model proposed in this paper that allowed for borrowing information from neighbors is important.  Additionally, because models for a single county struggled to converge, we also only fit these models to estimate a single change point; there would not be enough information to estimate very many (if any) change points beyond one.

Second, we found that for counties that did converge, estimates for change points and coefficients had very large uncertainty ranges and large overlap with other counties  further illustrating a need to combine the analysis for all counties.  We provide a more detailed examination into four counties here.

\begin{figure}
    \centering
    \includegraphics[width=0.52\textwidth,keepaspectratio]{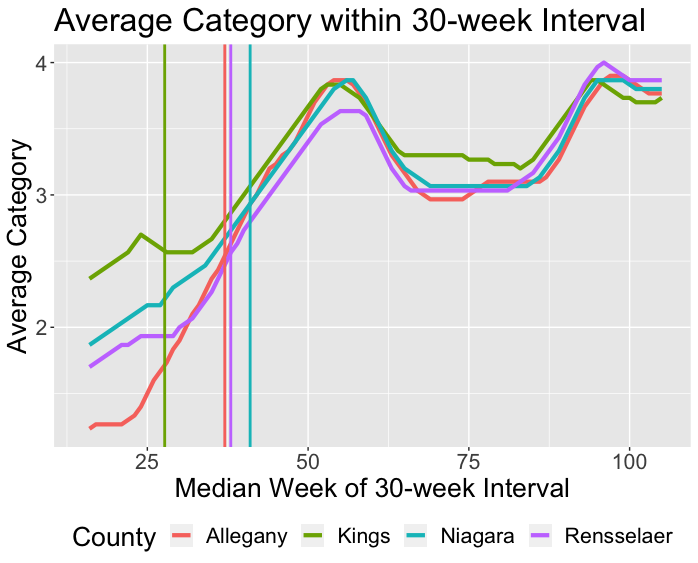}
    \caption{30-week average transmission levels for four counties in New York.  The vertical line in the corresponding color represents the posterior mean change point estimated when the model was fit to the single county data.}
    \label{fig:countycp}
\end{figure}

We explore four counties from across the state and with differing populations.  Specifically, Allegany County (southwest, rural, population 46091), Kings County (southeast, urban, population 2559903), Niagara County (northwest, urban/suburban, population 209281), and Rensselaer County (northeast, urban/suburban, population 158714).  The posterior mean change point for the four counties is listed in Table \ref{tab:countycp}.  Figure \ref{fig:countycp} provides a plot of the smoothed (30-week averages) transmission levels for the four counties with vertical lines representing the mean change point.  Notice that all four of these estimated change points are before any counties were administered vaccinations (which began around week 50 in New York, or December 28, 2020).  This makes sense that models would identify a change in relationships at this time; however, it's not useful in identifying and quantifying more practical changes.  Specifically, consider the coefficient estimates for vaccinations before the change point, $\beta_2$, in Table \ref{tab:countycp}.  The coefficients are very large both positively and negatively, particularly  relative to the probit link.  This is because before the change point there are no vaccinations so the posterior is simply returning the prior\footnote{Note that to avoid singularity in the $X$ matrix before the change point, we added a small amount of noise to the values of the covariates.}.  However, after the change point, once vaccines have started to be administered, the coefficients $\beta_2^*$ are much smaller, but uncertainty ranges still overlap with one another and include 0.  We see a similar behavior with the coefficients for previous weeks deaths, but there isn't a common time when this covariate remains constant for all counties.  These challenges indicate that a model that allows for borrowing information from neighbors is needed.

\begin{table}[!hbt]
\small
    \centering
        \caption{Posterior mean (and 95\% credible interval) for: the week of the change point, the corresponding date of the change point, and coefficients before (without $*$) and after (with $*$) the change point for four counties when the model was fit to the single county data.}
    \label{tab:countycp}
    \begin{tabular}{c|cc|c|c|c|c}
 \hline
       & \textbf{Change} &  & \multicolumn{2}{c|}{\textbf{Prev Week Death}} & \multicolumn{2}{c}{\textbf{Vaccination}}  \\
         \textbf{County}  &  \textbf{Point}  &  \textbf{Date} & $\beta_1$ & $\beta_1^*$ & $\beta_2$ &$\beta_2^*$\\\hline \hline
         Allegany & 38 & Sep 28 & 0.743 & 0.404 & 7715.6 & -2.97\\
          & (35, 38) & 2020 & (-0.11, 1.75) & (-0.02, 0.89) & (-1714, 17610) & (-9.65, 2.90)\\\hline
         Kings & 28 & Jul 27 & 37.686 & 0.369 & 2507.52 & 0.55\\
          & (15, 43) & 2020 & (0.62, 60.63) & (-0.29, 0.98) & (-23402, 45493) & (-3.73, 4.43)\\\hline
         Niagara & 41 & Oct 26 & 0.9 & 0.575 & -2028.81 & -5.137\\
         & (40, 42) & 2020  & (0.14, 2.07) &(0.08, 1.13) & (-11313, 7616) & (-12.92, 0.23)\\\hline
         Rensselaer & 37 & Sep 21 & 678.57 & 0.266 & -1395.54 & -1.833\\
         & (11, 44) & 2020 & (-6053, 15661) &(-0.11, 0.69) & (-12352, 7351) & (-8.52, 4.60)\\\hline
    \end{tabular}
\end{table}

\section{Posterior calculation}
\label{appendix-post}
This section presents the detailed calculations for the posterior distributions needed in the Gibbs sampler. The full joint posterior distribution can be written as
\begin{eqnarray*}
f(\bm\theta,\bm\theta^*,\bm{U},\bm{V},\bm{\pi},\bm{\phi},\bm{\delta},t_0|\bm{Y}) &\propto& f(\bm{Y}|\bm\theta,\bm\theta^*,\bm{U},\bm{V},\bm{\pi},\bm{\delta},\bm{\phi},t_0) f(\bm{\pi}|\bm{U},\bm{V},\bm\theta,\bm\theta^*,\bm{\delta},\bm{\phi},t_0) f(\bm{V}|\bm{\phi_v}) \\
&& f(\bm{U}|\bm{\phi}_u)f(\bm\theta|\omega_s) f(\bm\theta^*|\omega_s^*) f(t_0) f(\bm{\phi})f(\omega_s)f(\omega_s^*)f(\bm{\delta})\\
&\propto& f(\bm{Y}|\bm\theta,\bm\theta^*,\bm{U},\bm{V},\bm{\pi},\bm{\phi},t_0) f(\bm{\pi}^-|\bm{U}^-,\bm\theta,\bm\theta^*,t_0) \\
&& f(\bm{\pi}^+|\bm{U}^+,\bm{V}^+,\bm\theta^*,t_0)f(\bm{V}|\bm{\phi_v}) f(\bm{U}|\bm{\phi}_u)f(\bm\theta|\bm\omega_s)\\
&& f(\bm\theta^*|\bm\omega_s^*) f(t_0) f(\bm{\phi})f(\bm\omega_s)f(\bm \omega_s^*)f(\bm{\delta}),
\end{eqnarray*}
where $\bm{\phi}=(\bm{\phi}_u, \bm{\phi}_v)'$, $\bm{\phi}_u=(\phi_{us},\phi_{ut})'$, and $\bm{\phi}_v=(\phi_{vs},\phi_{vt})'$.

In order to calculate the conditional posterior distribution of $\bm{\theta}$, we need to take only the terms involving $\bm{\beta}$. On simplification, that yields
\begin{eqnarray*}
f(\bm\theta|\mathcal{F}) &\propto& f(\bm{\pi}^-|\bm{U}^-,\bm\theta,\bm\theta^*,t_0)f(\bm\theta|\omega_s) \\
&\propto& \exp\left\{  \frac{-1}{2} \left(  \bm{\theta}' \left[ \frac{X^{-'}X^-}{\sigma_{\epsilon}^2} \right]  \bm{\theta} -2\bm{\theta}'  \left[ \frac{X^{-'}(\bm{\pi}^- -\bm{U}^-)}{\sigma_{\epsilon}^2} \right]   \right)   \right\} \exp\left\{  \frac{-1}{2} \left(  \bm{\theta}' \left[\frac{\Psi^{-1}}{k} \right]  \bm{\theta}   \right)   \right\}\\
&\propto& \exp\left\{  \frac{-1}{2} \left(  \bm{\theta}' \left[ \frac{X^{-'}X^-}{\sigma_{\epsilon}^2} + \frac{\Psi^{-1}}{k} \right]  \bm{\theta} -2\bm{\theta}'  \left[ \frac{X^{-'}(\bm{\pi}^- -\bm{U}^-)}{\sigma_{\epsilon}^2} \right]   \right)   \right\},
\end{eqnarray*}
where $\Psi$ are block diagonal matrices with entries in the order $(\I_G,\Omega_{s_1},\Omega_{s_2},\hdots,\Omega_{s_H})$.

Comparing the above with multivariate normal distribution, we get,
\begin{equation}
\label{eq:appendix-beta-posterior}
     \bm{\theta}| \mathcal{F} \sim N_k\left( \Sigma_{\bm{\theta}}  \left[ \frac{X^{-'}(\bm{\pi}^{-}-\bm{U}^- )}{\sigma_{\epsilon}^2}   \right], \Sigma_{\bm{\theta}}  \right),
\end{equation}
where dispersion matrices $\Sigma_{\bm{\theta}}$ is defined as
\begin{equation*}
        \Sigma_{\bm{\theta}} =\left[ \frac{X^{-'}X^-}{\sigma_{\epsilon}^2}+ \frac{\Psi^{-1}}{k}\right]^{-1}.
\end{equation*}

Similarly, we obtain the conditional posterior for $\bm{\theta}^*$.

For the conditional posterior of the vector $\bm{\pi}$, we consider the $i^{th}$ element of it (denoted by $\pi_i$, as mentioned before) and similarly for vectors $\bm{U}$, $\bm{V}$ and $\bm{Y}$. Since all the elements of $\bm{\pi}$ vector will be independent of each other conditional on $\bm{U}$ and $\bm{V}$ vectors, we can write
\begin{equation*}
    f(\pi_i|\mathcal{F}) \propto
    \begin{cases}
        f(\pi_i|u_i,\bm{\theta})\, I(\delta_{j-1}<\pi_i<\delta_{j}) , & \text{for } t(i) \leqslant t_0,y_i=j, \\
        f(\pi_i|u_i,v_i,\bm{\theta}^*)\, I(\delta_{j-1}<\pi_i<\delta_{j}) , & \text{for } t(i) > t_0,y_i=j.
    \end{cases}
\end{equation*}

Thus we can write the conditional posterior for $\pi_i$ as
\begin{equation}
\label{eq:apdx-sigma_pi_ij-posterior-dist-ord}
\pi_{i}|\mathcal{F} \sim 
    \begin{cases} 
    TN(x'_i\bm{\theta} + u_i ,\sigma_{\epsilon}^2;\delta_{j-1},\delta_{j}), & \text{for } t(i) \leqslant t_0,y_i=j, \\
    TN(x'_i\bm{\theta}^*+u_i + v_{i},\sigma_{\epsilon^*}^2;\delta_{j-1},\delta_{j}), & \text{for } t(i) > t_0,y_i=j,
    \end{cases}
\end{equation}
where $TN(a,b;c,d)$ denotes the univariate truncated normal distribution, obtained by truncating a normal distribution with mean $a$ and variance $b$ in the interval $(c,d)$.

Next, since $\delta_j$ should be between $\delta_{j-1}$ and $\delta_{j+1}$, the conditional posterior distribution for $\delta_j$ is written as
\begin{eqnarray*}
    f(\delta_j|\mathcal{F}) &\propto& f(\bm{Y}|\bm\theta,\bm\theta^*,\bm{U},\bm{V},\bm{\pi},\bm{\delta},\bm{\phi},t_0) f(\bm\delta)\\
    &\propto& I(\delta_{j-1}<\delta_j<\delta_{j+1})\intop_{\mathcal{A}_j} \bigg( \exp{\bigg\{ \frac{-1}{2\sigma_{\epsilon}^2} \norm{\bm{\pi}^- -X^-\bm{\theta} - \bm{U}^- }^2 \bigg \} } {\mathrm{d} \bm{\pi}^-} \\
    && \intop_{\mathcal{B}_j} 
     \exp{\bigg\{ \frac{-1}{2\sigma_{\epsilon^*}^2}\norm{\bm{\pi}^+-X^+\bm{\theta}^* - \bm{U}^+- \bm{V}^+}^2 \bigg \} } {\mathrm{d} \bm{\pi}^+},
\end{eqnarray*}
where $\mathcal{A}_j$ and $\mathcal{B}_j$ denote the integration for all the observations in the $j^{th}$ category with bounds for integration are $\delta_{j-1}$ and $\delta_{j}$ and $(j+1)^{th}$ category with bounds for integration are $\delta_{j}$ and $\delta_{j+1}$ before and after the changepoint $t_0$, respectively.

Now, let $\bm{\pi}_j^-$ and $\bm{\pi}_{j+1}^-$ be the vector that represents only those values from vector $\bm{\pi}^-$ where the corresponding categories for the vector $\bm{Y}^-$ are $j$ and $j+1$, respectively. Similarly, $X_j^-$, $\bm{U}_j^-$, $X_{j+1}^-$, and $\bm{U}_{j+1}^-$ are defined. Also, let $\bm{\pi}_j^+$ and $\bm{\pi}_{j+1}^+$ be the vector that represents only those values from vector $\bm{\pi}^+$ where the corresponding categories for the vector $\bm{Y}^+$ are $j$ and $j+1$, respectively. Similarly, $X_j^+$, $\bm{U}_j^+$, $\bm{V}_j^+$, $X_{j+1}^+$, $\bm{U}_{j+1}^+$, and $\bm{V}_{j+1}^+$ are defined.
\begin{equation}
\label{eq:apdx-delta-post3}
\small
\begin{split}
    f(\delta_j|\mathcal{F}) & \propto I(\delta_{j-1}<\delta_j<\delta_{j+1})\intop_{\delta_{j-1}}^{\delta_{j}}  \exp{\bigg\{ \frac{-1}{2\sigma_{\epsilon}^2} \norm{\bm{\pi}_j^- -X_j^-\bm{\theta} - \bm{U}_j^- }^2 \bigg \} } {\mathrm{d} \bm{\pi}_j^-} \\
    &  \intop_{\delta_{j}}^{\delta_{j+1}}  \exp{\bigg\{ \frac{-1}{2\sigma_{\epsilon}^2} \norm{\bm{\pi}_{j+1}^- -X_{j+1}^-\bm{\theta} - \bm{U}_{j+1}^- }^2 \bigg \} } {\mathrm{d} \bm{\pi}_{j+1}^-} \\
    &\intop_{\delta_{j-1}}^{\delta_{j}} 
     \exp{\bigg\{ \frac{-1}{2\sigma_{\epsilon^*}^2}\norm{\bm{\pi}_j^+-X_j^+\bm{\theta}^* - \bm{U}_j^+- \bm{V}_j^+}^2 \bigg \} } {\mathrm{d} \bm{\pi}_j^+}\\
    &\intop_{\delta_{j}}^{\delta_{j+1}} 
     \exp{\bigg\{ \frac{-1}{2\sigma_{\epsilon^*}^2}\norm{\bm{\pi}_{j+1}^+-X_{j+1}^+\bm{\theta}^* - \bm{U}_{j+1}^+- \bm{V}_{j+1}^+}^2 \bigg \} } {\mathrm{d} \bm{\pi}_{j+1}^+},
\end{split}
\end{equation}

Now, we calculate the conditional posterior for the decay parameter $\phi_{vt}$, for other decay parameters one can similarly compute the conditional posterior. The decay parameter values are assumed to be between 0 and 3 as mentioned in the main paper. This comes from the relationship that for unit distance in space or time, the decay parameter being 3 or more translates to the correlation being negligible. Using this prior, the conditional posterior of $\phi_{vt}$ can be written as
\begin{equation}
\label{eq:apdx-phi-vt}
\begin{split}
    f(\phi_{vt}|\mathcal{F}) &\propto f(\bm{V}|\phi_{vt})f(\phi_{vt})\\
    &\propto |\Sigma_{vt}|^{-n/2} \exp \Bigg\{\frac{- \bm{V}'(\Sigma_{vt}^{-1} \otimes \Sigma_{vs}^{-1}) \bm{V}}{2\sigma_{v}^2}\Bigg\} I(0<\phi_{vt}<3).
\end{split}
\end{equation}

As discussed in Section 3.2 of the main paper, we divide the vector $\bm{V}$ into $\bm{V}_1,\bm{V}_2,\hdots,\bm{V}_n$ with $\bm{V}_t$ being the $t^{th}$ column of the $V$ matrix in eq. (3.4) of the main paper.
Hence,
\begin{equation*}
    \bm{V}_t|\bm{V}_{-t} \sim N_T (\bm\mu_{vc},\sigma_v^2\Sigma_{vc}),
\end{equation*}
where $\bm\mu_{vc} = (\Sigma_{12}\Sigma_{22}^{-1} \otimes \I_n)\bm{V}_{-t}$ and $\Sigma_{vc}=(\Sigma_{11} -\Sigma_{12}\Sigma_{22}^{-1} \Sigma_{21})\otimes \Sigma_{vs}$. Then, for the conditional posterior distribution of $\bm{V}_t$, if $t^{th}$ time point is greater than $t_0$, we note that
\begin{equation*}
    f(\bm{V}_t|\mathcal{F}) \propto
   f(\bm{\pi}^+|\bm{U}^+,\bm{V}^+,\bm\theta^*,t_0)    f(\bm{V}_t|\bm{V}_{-t}),
\end{equation*}
thereby implying that
\begin{eqnarray*}
 f(\bm{V}_t|\mathcal{F})   &\propto& \exp \left \{-\frac{1}{2\sigma_{\epsilon^*}^2} \norm{\bm{\pi}^+ - X^+\bm{\theta}^* - \bm{U}^+ - \bm{V}^+}^2\right \} \\
 && \exp \left \{-\frac{1}{2\sigma_v^2} (\bm{V_t}  - \bm{\mu}_{vc})'\Sigma_{vc}^{-1}(\bm{V}_t  - \bm{\mu}_{vc})\right \}.
\end{eqnarray*}

Taking only the terms involving $\bm{V}_i$, the above can be shown to be proportional to
\begin{eqnarray*}
f(\bm{V}_t|\mathcal{F})&\propto& \exp  \left \{-\frac{1}{2\sigma_{\epsilon^*}^2} (\bm{V}_t'\bm{V}_t -2\bm{V}_i'(\bm{\pi}_t - X_t\bm{\theta}^* - \bm{U}_t)) -\frac{1}{2\sigma_v^2}  (\bm{V}_t'\Sigma_{vc}^{-1}\bm{V}_t -2\bm{V}_t'\Sigma_{vc}^{-1}\bm{\mu}_{vc})\right \} \\
&\propto& \exp  \left \{-\frac{1}{2}  \biggl(\bm{V}_t'       \biggl(\frac{\Sigma_{vc}^{-1}}{\sigma_v^2} + \frac{\I_n}{\sigma_{\epsilon^*}^2} \biggr)  \bm{V}_t -2\bm{V}_t'\left(\frac{\Sigma_{vc}^{-1}\bm\mu_{vc}}{\sigma_v^2} + \frac{\bm{\pi}_t - X_t\bm\theta^* - \bm{U}_t}{\sigma_{\epsilon^*}^2}\right)\biggr)        \right \}
\end{eqnarray*}

Straightforward calculation through multivariate normal theory and considering the other case when $i^{th}$ time point is less than or equal to $t_0$ and doing similar computations for $\bm{U}$ gives
\begin{equation}
\label{eq:apdx-v_i-posterior-ord}
\bm{V}_{t}| \mathcal{F} \sim
 \begin{cases}
    N_n (\mu_{vc},\sigma_{v}^2 \Sigma_{vc}
    ), & \text{for } t \leqslant t_0,\\
    N_n \biggl( \Sigma_{V_t}\left(\frac{\Sigma_{vc}^{-1}\bm\mu_{vc}}{\sigma_{v}^2} + \frac{\bm{\pi}_{t} - X_{t}\bm{\theta}^* - \bm{U}_t}{\sigma_{\epsilon^*}^2}\right),
    \Sigma_{V_t}\biggr), & \text{for } t > t_0,
    \end{cases}
\end{equation}

\begin{equation}
\label{eq:apdx-u_i-posterior-ord}
\bm{U}_{t}| \mathcal{F} \sim
 \begin{cases}
    N_n \biggl( \Sigma_{U_t}\left(\frac{\Sigma_{uc}^{-1}\bm\mu_{uc}}{\sigma_{u}^2} + \frac{\bm{\pi}_{t} - X_{i}\bm{\theta}}{\sigma_{\epsilon}^2}\right),
           \Sigma_{U_t}\biggr), & \text{for } t \leqslant t_0,\\
    N_n \biggl( \Sigma_{U_t}\left(\frac{\Sigma_{uc}^{-1}\bm\mu_{uc}}{\sigma_{u}^2} + \frac{\bm{\pi}_{t} - X_{t}\bm{\theta}^* - \bm{V}_t}{\sigma_{\epsilon}^2}\right),
           \Sigma_{U_t}\biggr), & \text{for } t > t_0,
    \end{cases}
\end{equation}
where $\Sigma_{V_t}$ and $\Sigma_{U_t}$ are defined as
\begin{equation*}
    \Sigma_{V_t} = \left[\frac{\Sigma_{vc}^{-1}}{\sigma_{v}^2}+\frac{\I_n}{\sigma_{\epsilon^*}^2}\right]^{-1}, \;
    \Sigma_{U_t} = \left[\frac{\Sigma_{uc}^{-1}}{\sigma_{u}^2}+\frac{\I_n}{\sigma_{\epsilon^*}^2}\right]^{-1}.
\end{equation*}

Finally, considering the set $S_T=\{0, 1, 2, \hdots, T\}$, following the steps described in the main manuscript, we obtain the conditional posterior for $t_0$ as
\begin{equation}
\label{eq:apdx-t_0-post}
\begin{split}
    f(t_0|\mathcal{F}) &\propto f(\bm{\pi}^-|\bm{U}^-,\bm\theta,\bm\theta^*,t_0) f(\bm{\pi}^+|\bm{U}^+,\bm{V}^+,\bm\theta^*,t_0) f(t_0)\\
    &\propto I(t_0 \in S_T) (2\pi \sigma_{\epsilon}^2)^{\frac{-nT_{t_0}^-}{2}}\exp \bigg \{ \frac{-1}{2\sigma_{\epsilon}^2} \norm{\bm{\pi}  ^{-}-(X^-\bm{\theta}+\bm{U}^-)}^2 \bigg \} \\
    & (2\pi \sigma_{\epsilon^*}^2)^{\frac{-nT_{t_0}^+}{2}}\exp \bigg \{ \frac{-1}{2\sigma_{\epsilon^*}^2} \norm{\bm{\pi}^{+}-(X^+\bm{\theta}^*+\bm{U}^++\bm{V}^+)}^2 \bigg \}.
\end{split}
\end{equation}


\end{document}